\documentclass[review]{elsarticle}
\usepackage{xcolor}
\usepackage{lineno,hyperref}
\usepackage{amsmath}
\usepackage{subcaption}
\journal{Journal of \LaTeX\ Templates}

\begin{document}
\begin{frontmatter}

\title{Neuronal coupling benefits the encoding of weak periodic signals in symbolic spike patterns}

\author[1]{Maria Masoliver}
\ead{maria.masoliver@upc.edu}
\author[2]{Cristina Masoller\corref{cor1}}
\ead{cristina.masoller@upc.edu}

\cortext[cor1]{Corresponding author}
\fntext[fn1]{Universitat Polit\`{e}cnica de Catalunya. Departament de F\'{i}sica.\\ 
Edifici Gaia, Rambla de Sant Nebridi 22, 08222, Terrassa, Barcelona, Spain.}

\begin{abstract}
The biophysical mechanisms by which an input signal elicits a neuronal response are well known (sufficiently large inputs change the membrane potential of the neuron and generate electrical pulses, known as action potentials or spikes), yet, a good understanding of how neurons use these spikes to encode the signal information remains elusive. Recent theoretical studies have focused on how neurons encode a weak periodic signal (that by itself is unable to generate spikes) in a noisy environment, where stochastic electrical fluctuations that do not encode any information occur. Analyzing spike sequences generated by individual neurons and by two coupled neurons (that were simulated with the stochastic FitzHugh-Nagumo model), it has been found that the relative timing of the spikes can encode the signal information. Using a symbolic method to analyze the spike sequence, preferred and infrequent spike patterns were detected, whose probabilities vary with both, the amplitude and the frequency of the signal. To investigate if this encoding mechanism is plausible also for neuronal ensembles, here we analyze the activity of a group of neurons, when they all perceive a weak periodic signal.  We find that, as in the case of one or two coupled neurons, the probabilities of the spike patterns, now computed from the spike sequences of all the neurons, depend on the signal's amplitude and period, and thus, the patterns' probabilities encode the information of the signal.  We also find that the resonances with the period of the signal or with the noise level are more pronounced when a group of neurons perceive the signal, in comparison with when only one or two coupled neurons perceive it. Neuronal coupling is beneficial for signal encoding as a group of neurons is able to encode a small-amplitude signal, which could not be encoded when it is perceived by just one or two coupled neurons.  Interestingly, we find that for a group of neurons, just a few connections with one another can significantly improve the encoding of small-amplitude signals. Our findings indicate that information encoding in preferred and infrequent spike patterns is a plausible mechanism that can be employed by neuronal populations to encode weak periodic inputs, exploiting the presence of neural noise.
\end{abstract}


\end{frontmatter}

\section{Introduction}
A mechanical input such as tapping someone's knee elicits a stretch reflex as a response. The biophysical mechanism is known, the muscle stretches as a consequence of the tapping to the tendon, which triggers the generation of spikes by a sensory neuron, which in turn triggers the generation of spikes by a motor neuron, leading to muscle contraction and causing the lower leg to bounce back \citep{PEA00,NOL02,KNI14}. On the other hand, the signal encoding mechanism is also known, the frequency at which the sensory neuron fires encodes the information about how fast the muscle is stretching \citep{HUN90}. In turn, the firing rate of the motor neuron encodes the information about the muscle force when it contracts \citep{MON77}. This is an example of a neural circuit (two neurons interconnected by a synapse), which uses the firing rate code as a coding scheme for an external input. Yet, neurons encode information of different types of signals using different encoding mechanisms, which are not yet fully understood. Neurons can represent external or internal inputs in the timing of the individual spikes, in the relative timing of the spikes of two or more neurons, in the individual firing rate, in the average firing rate of a population of neurons, in the spike arrival times, among other coding schemes \citep{KNI72, CAR96, SAK99, MAS02, ARA04, NEL05, AVE06, QUI09, WAN10,KUM10,TCH11, KNI14,CAR18,LAZ18}. 

Understanding the neural code is crucial, not only to gain knowledge of the operation of the central nervous system, but also, to advance artificial intelligence systems based on neural networks that use spike-processing operations for classification, pattern recognition, logic operations, etc. \citep{PRU16, SHA16,SHE17,book}. 

Efforts have focused on understanding the role, on neural coding, of neural noise (stochastic electrical fluctuations that do not encode any information \cite{DON11}) and  spike temporal correlations, in particular, for encoding and processing weak sensory signals. By analyzing the coefficient of variation of the inter-spike interval (ISI) distribution (the standard deviation of the ISI distribution divided by the mean), the well-known phenomena of stochastic resonance and coherence resonance have been found. While stochastic resonance \citep{LIN95, YIL13} refers to the enhancement of weak signal detection, coherence resonance \citep{PIK97, ZHO01, KWO02, KWO05, Bal2014, MAS17} refers to the regularization of the spike train, for an optimal level of noise. On the other hand, ISI correlations lasting several ISIs have been studied by using the lagged serial correlation coefficient, which measures linear relations between
sequential interspike intervals \citep{NEI05, NEI11,NES10,AKE11, SCH13,BRA19}.

An alternative technique, known as {\it ordinal analysis} \citep{BAN02,AMIGO} has also been used to detect nonlinear ISI correlations. In general terms, ordinal analysis transforms a time series into a sequence of symbols, known as ordinal patterns, considering the temporal order relations among the data points in the time series. A main advantage of the ordinal symbolic approach is that it provides a straightforward way to quantify how much information is contained in a sequence of spikes (i.e., to apply Information Theory to the study of the neural code \cite{STR98,BOR99}): by  counting the number of times each ordinal pattern appears in the spike sequence, the probabilities of the different patterns can be estimated, providing a quantification of the information content.  Ordinal analysis has been widely used to investigate biomedical signals, for example, to quantify directionality of coupling in cardio-respiratory data \cite{BAH08}, to characterize neuronal spike trains \cite{LI11,rosso1,rosso2}, to distinguish healthy subjects from patients suffering from congestive heart failure \citep{PAR12}, to classify neurophysiological data \cite{CAO04,ARR13, QUI18, ECH19}, etc.

In order to understand how a single neuron can encode a weak periodic input signal (that by itself is unable to generate spikes) in the presence of neural noise, Aparicio Reinoso et al. \cite{REI16} have applied ordinal analysis to spike sequences simulated with the stochastic FitzHugh-Nagumo (FHN) model  \cite{FIT61, NAG62}.  It was found that the periodic signal induces temporal order in the form of more and less expressed ordinal patterns. The probabilities of the patterns encode the signal information, as they depend on both, the amplitude and the period of the signal.  In a follow up study \cite{MAS18}, the role of a second neuron that does not perceive the weak signal was analyzed. It was found that the signal is still encoded in the form of preferred and infrequent ordinal patterns, whose probabilities again depend on the period and amplitude of the signal.

An open question is whether this encoding mechanism can be employed by a population of neurons. To answer this question, here we use the stochastic FHN model to simulate the activity of a group of neurons, when they all perceive a periodic signal that is weak enough such that by itself (in the absence of noise) it is unable to generate spikes. Thus, as in previous studies, the neuronal ensemble encodes the signal in spikes sequences which are generated due to the interplay of the signal and the noise.  Our main findings can be summarized as follows: (i) the ensemble is able to encode lower amplitude signals, in comparison with the signal amplitude that can be encoded by a single neuron or by two coupled neurons; (ii) the noise-induced and period-induced resonances (some ordinal patterns probabilities are minimum or maximum for particular values of the noise strength or signal period) observed in one \citep{REI16} or two coupled neuron \citep{MAS18} become more pronounced for the neuronal ensemble and (iii) just a few connections among the neurons can significantly improve the signal encoding.

This paper is organized as follows. Section~\ref{Model} presents the model equations, Sec.~\ref{Methods} presents the ordinal analysis method, Sec.~\ref{Results} presents the results and Sec.~\ref{Conclusions} presents the conclusions.

\section{Model}
\label{Model}
The FitzHugh-Nagumo model is one of the simplest (and yet quite realistic) models that describe excitable systems \cite{FIT61, NAG62,ACE04}. The equations describing the dynamics of an ensemble of coupled neurons are:
\begin{equation}
\begin{gathered}
	\epsilon \dot{u_i}= u_i - \frac{u_i^3}{3} - v_i  + a_0\cos(2\pi t/T) +  \frac{\sigma}{k_i } \sum_j^N{a_{ij}(u_j - u_i)} +\sqrt{2D}\xi_i(t), \hspace{1 cm} i \neq j \\
	\dot{v_i} = u_i+ a.
	\end{gathered}
	\label{eq:model_2}
\end{equation}	
Here $N$ refers to the number of neurons, $v$ is known as the inhibitor variable and $u$ is known as the activator variable that represents the evolution of the membrane potential: in the excitable regime, if there is no external perturbation or it is not strong enough to overcome the threshold, the membrane potential is held at the resting potential (i.e., stable fixed point) whereas when there is an external perturbation strong enough to overcome the threshold, the membrane potential performs a spike (i.e., an action potential). Typical parameters in the excitable regime are  $a = 1.05$ and $\epsilon = 0.01$.  

The parameters $a_0$ and $T$ represent the amplitude and period of an external sinusoidal input, and are chosen such that the signal is sub-threshold: without noise the neurons do not fire spikes. $D\xi_i(t)$ represents an stochastic term of strength $D$, which is taken as Gaussian distributed, uncorrelated temporally and across the neuronal ensemble: $\langle \xi_i(t)\xi_j(t')\rangle =\delta_{ij} \delta(t-t')$ with $\langle \xi_i(t)\rangle = 0$ and  $\langle \xi_i^2(t)\rangle = 1$.

The neurons are mutually coupled with gap-junction connections, characterized by symmetric links ($a_{ij}=a_{ji}=1$ if neurons $i$ and $j$ are connected, else $a_{ij}=a_{ji}=0$). The coupling strength of each link is $\sigma$; to keep the total coupling strength uniform for all neurons, it is normalized by number of connections, $k_i=\sum_j a_{ij}$.  Regarding the coupling topology, we focus on all-to-all coupling (in this case $k_i = N-1$ for all $i$), but we also consider random connections. This allows us to analyze the influence of the number of links, as neurons $i$ and $j$ are connected with probability $p$ that is varied between 0 and 1. It will be interesting, for future work, to investigate more realistic topologies with, for example, modular or hierarchical structures.

The model equations are simulated, from random initial conditions, using the Euler-Maruyama method with an integration step of $dt = 10^{-3}$. For each set of parameters, the voltage-like variable of each neuron $u_i$ is analyzed and the sequence of inter-spike-intervals (ISIs) is computed, $\{{I_j}_i; {I_j}_i = {(t_{j+1} - t_j)}_i\}$ with $t_j$ defined by the condition $u_i(t_j) = 0$, considering only the ascensions.

\section{Ordinal analysis}
\label{Methods}
The ordinal method \cite{BAN02} is used to analyze each ISI sequence.  From the sequence $\{I_1, \dots I_i, \dots I_N\}$ (for clarity, the subindex that labels the neuron is removed) symbols known as ordinal patterns are obtained by comparing $D$ consecutive ISIs, based on their temporal relation.
For example, if we set $D = 2$, the total number of possible ordinal patterns is two: 01, for $I_1 < I_2$ and 10, for $I_1 > I_2$, while if we set $D = 3$, we have $3! = 6$ possible ordinal patterns: 012 ($I_3 > I_2 > I_1$), 021 ($I_2 > I_3 > I_1$), 102 ($I_3 > I_1 > I_2$), 120 ($I_2 > I_1 > I_3$), 201 ($I_1 > I_3 > I_2$) and 210 ($I_1 > I_2 > I_3$).  The number of possible ordinal patterns (i.e., the number of possible temporal relations) is determined by the number of permutations, $D!$.

Using the function defined in \cite{PAR12} the sequence of ordinal patterns is computed. In order to determine if there are some preferred/infrequent patterns in the ISI sequences, ordinal patterns probabilities are calculated, taking together all the ISI sequences. The ordinal probabilities are estimated as $p_i$ = $C_i/M$, where $C_i$ refers to the number of times the $i-$th pattern appears and $M = \sum_{i = 1}^{D!} C_i$ is the total number of ordinal patterns.  If ordinal patterns are equi-probable it does not exist a preferred order relation among the timing of spikes. Yet, if there are preferred/infrequent ordinal patterns, a non-uniform probability distribution is obtained. In order to distinguish between these two cases (uniform vs. non-uniform ordinal distribution) a binomial test is used: if all the ordinal patterns are within the interval $[p - 3\sigma_p, p + 3 \sigma_p]$ with $p = 1/D!$ and $\sigma_p = \sqrt{p(1-p)/M}$ the ordinal probabilities are consistent with the uniform distribution with 99.74$\%$ confidence level; else, some patterns are over or less expressed than others, and there is some degree of temporal order in the timing of the spikes. 

A large number of spikes are needed to precisely estimate the ordinal probabilities (the data requirements for a single neuron were analyzed in \cite{REI16}, see Fig. 6). As long simulations are computationally demanding, here we limit to consider ensembles of up to 50 neurons. We have analyzed the role of the number of neurons, and we expect that our findings will hold for larger ensembles. The simulations are done for a time long enough to obtain a total number of $10^5$ spikes. As in \cite{REI16,MAS18}, we use $D=3$. This choice is motivated by the fact that only short ISI correlations are expected since the spikes are noise-induced (the signal by itself does not induce spikes). 

\section{Results}
\label{Results}
\begin{figure}[!ht]
\centering
 \includegraphics[width=.6\linewidth]{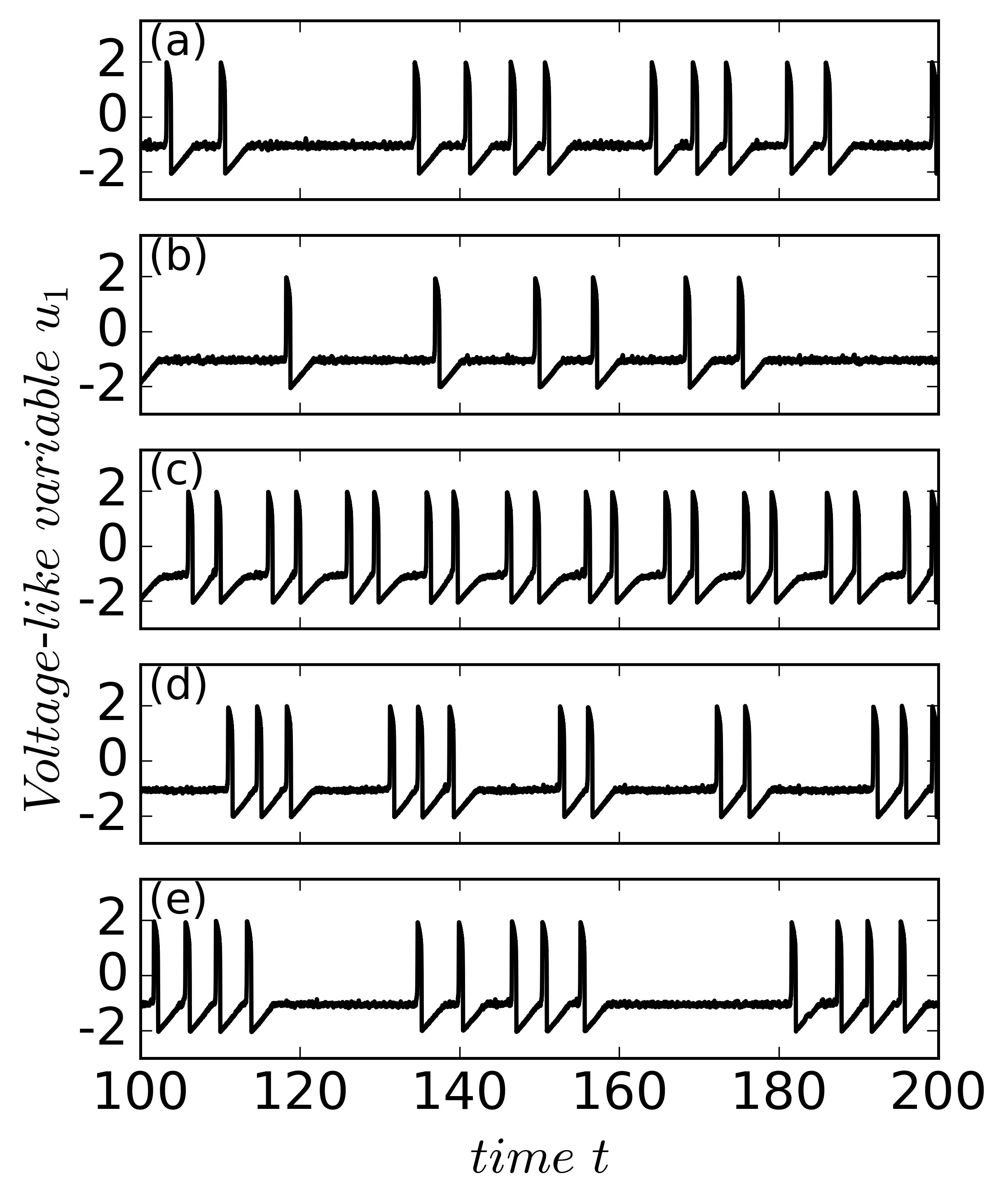}
\caption{Spiking activity of a neuron when no signal is applied ($a_0 = 0$) and the neuron (a) is uncoupled ($\sigma = 0$), (b) is coupled to a group of 50 neurons (all to all coupling, $\sigma = 0.05$). Activity of the neuron when it is coupled and a sinusoidal signal of amplitude $a_0 = 0.1$ and period (c) $T = 10$, (d)  $T = 20$, (e)  $T = 40$ is applied. The noise level is $D = 2.5\cdot 10 ^{-6}$. }
\label{Fig0}
\end{figure}

\begin{figure}[!ht]
\centering
\includegraphics[width=.3\textwidth]{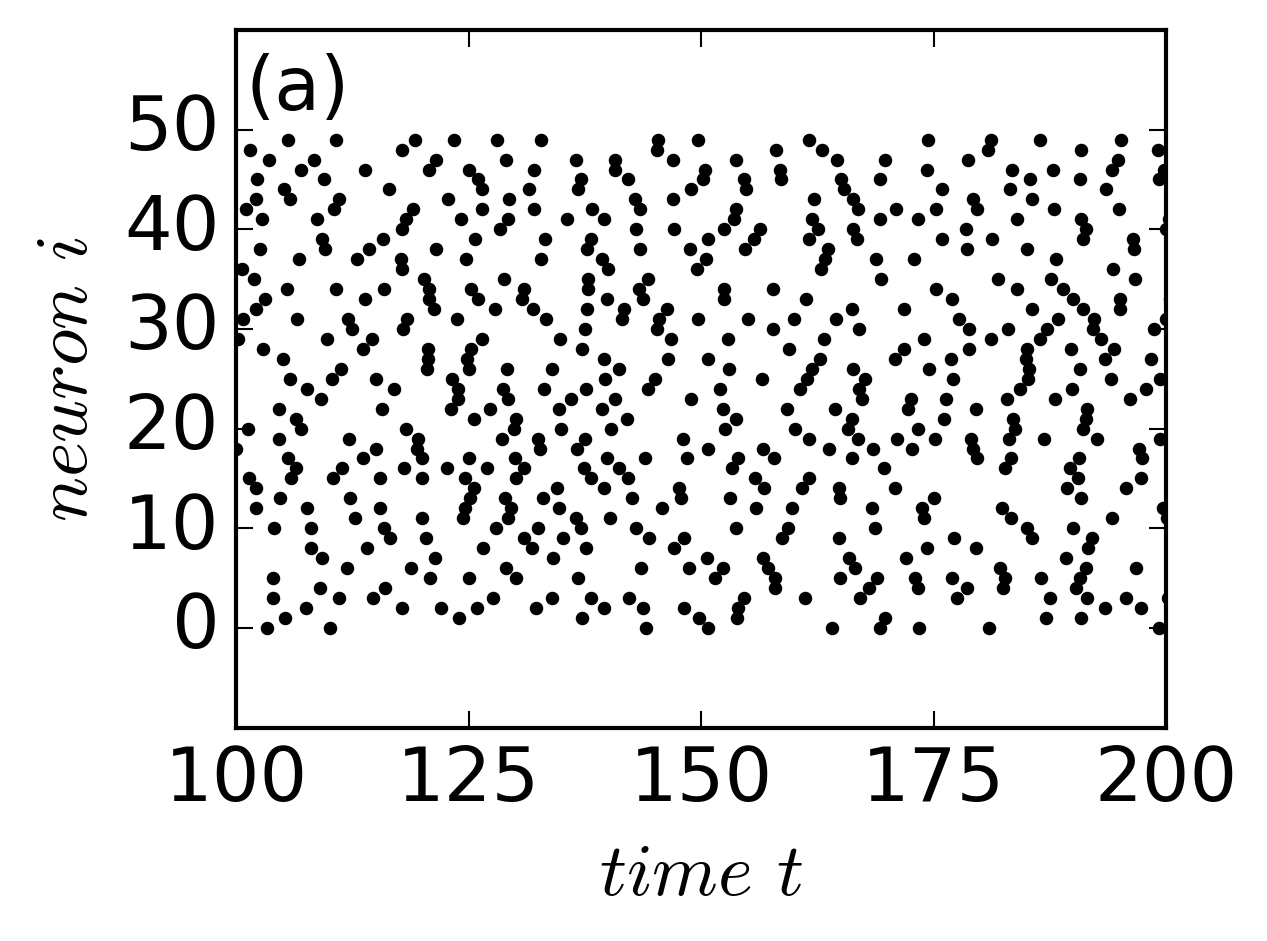}\quad
\includegraphics[width=.3\textwidth]{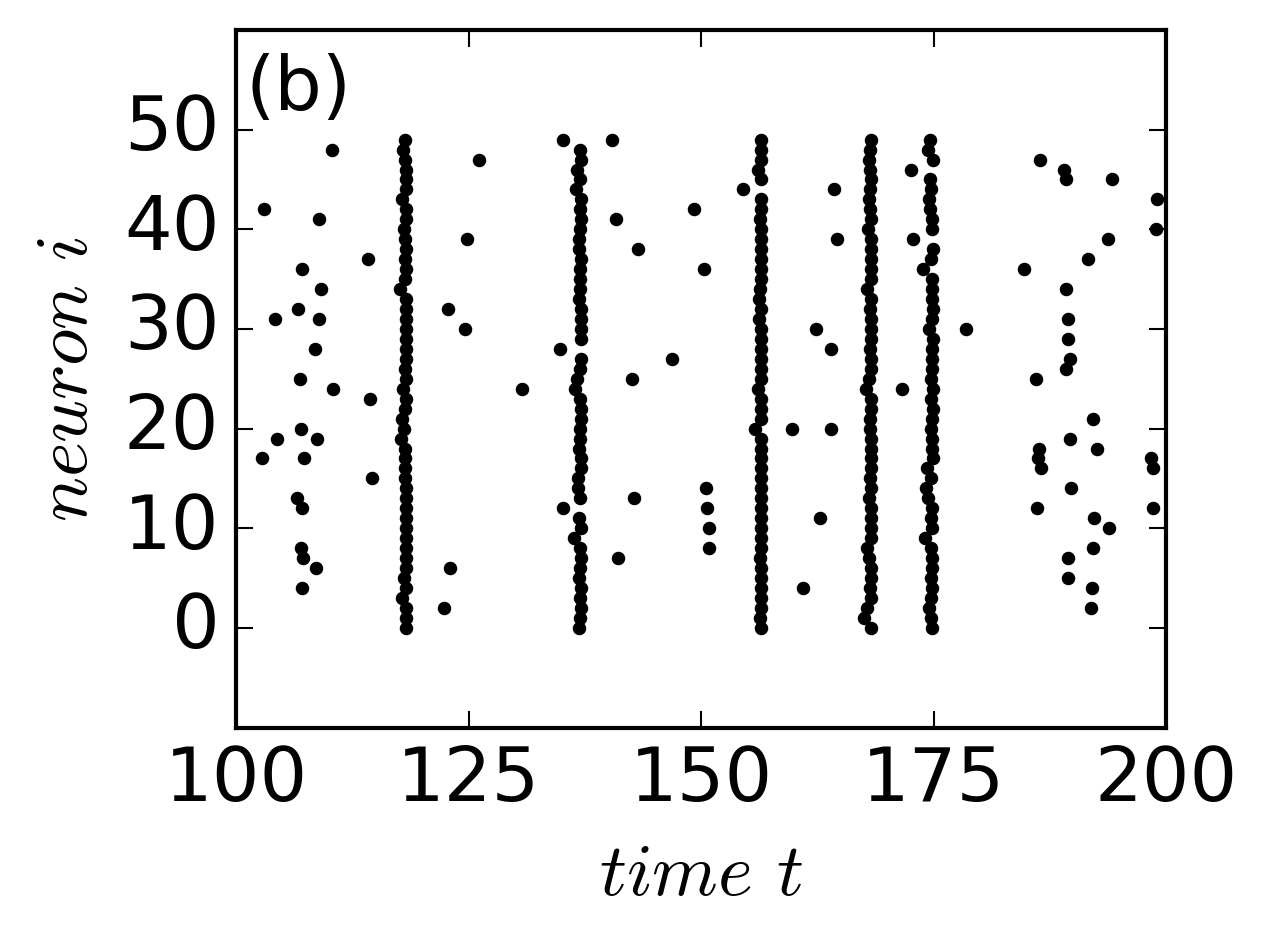}
\medskip
\includegraphics[width=.3\textwidth]{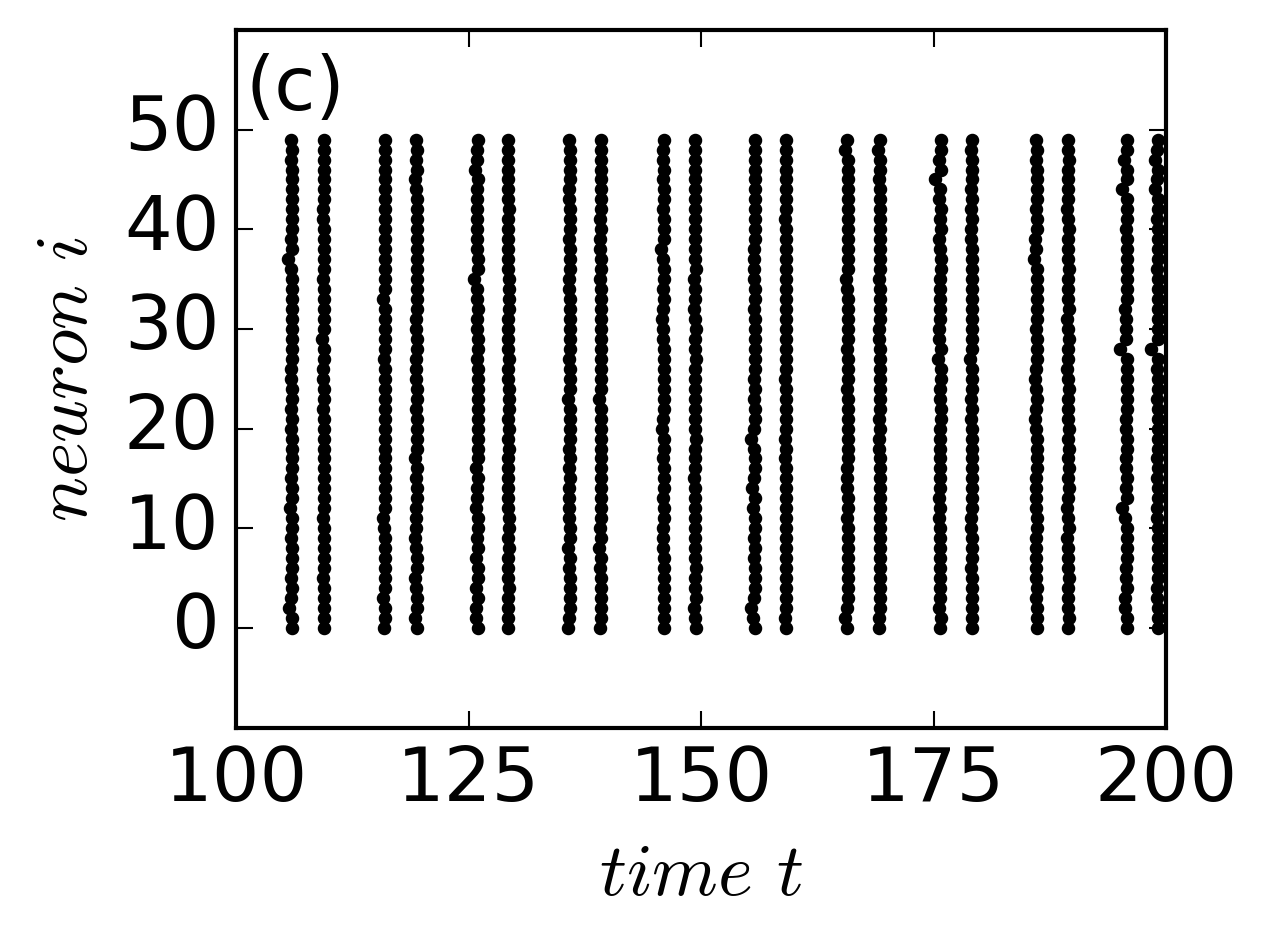}\quad
\includegraphics[width=.3\textwidth]{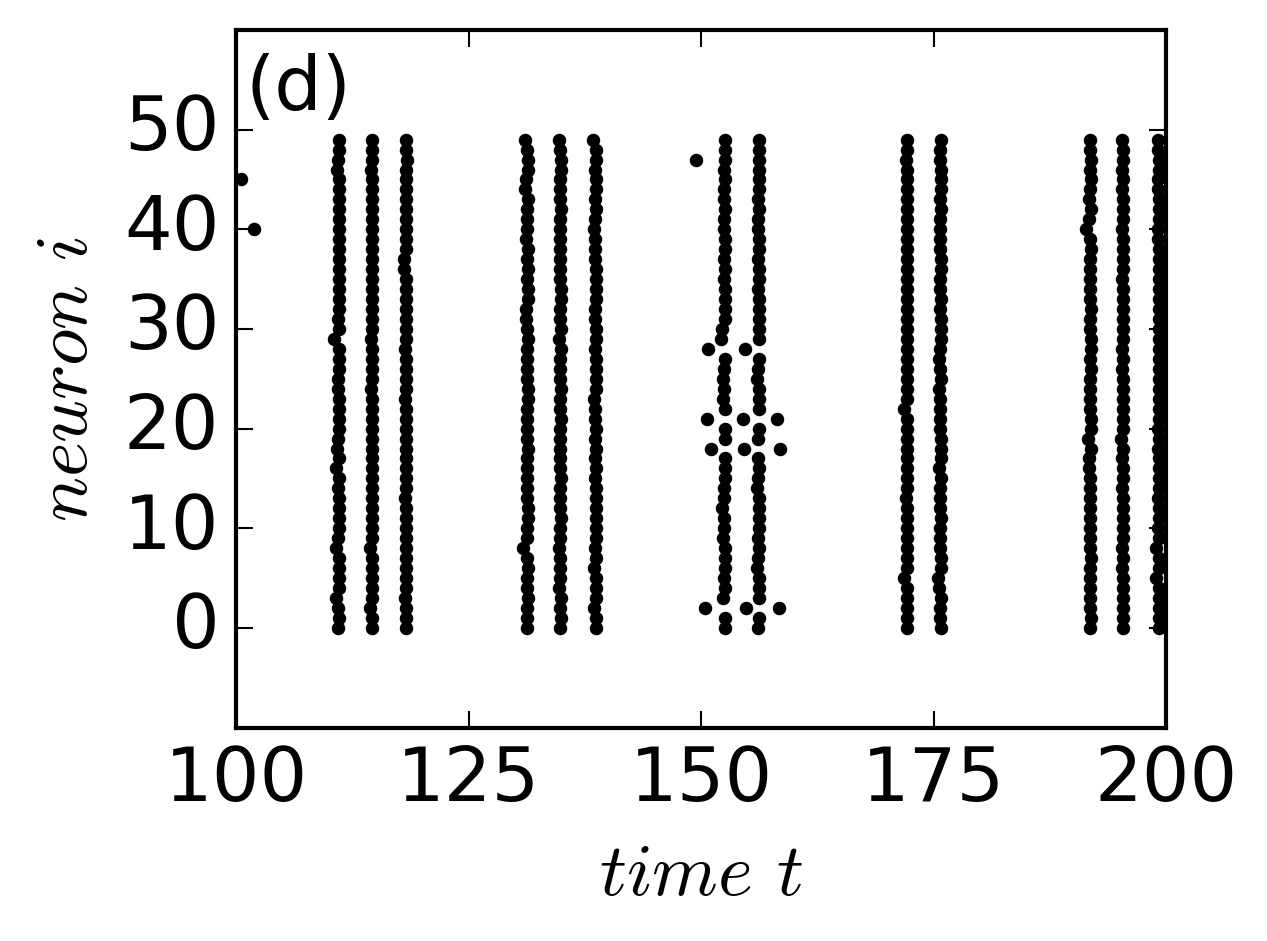}\quad
\includegraphics[width=.3\textwidth]{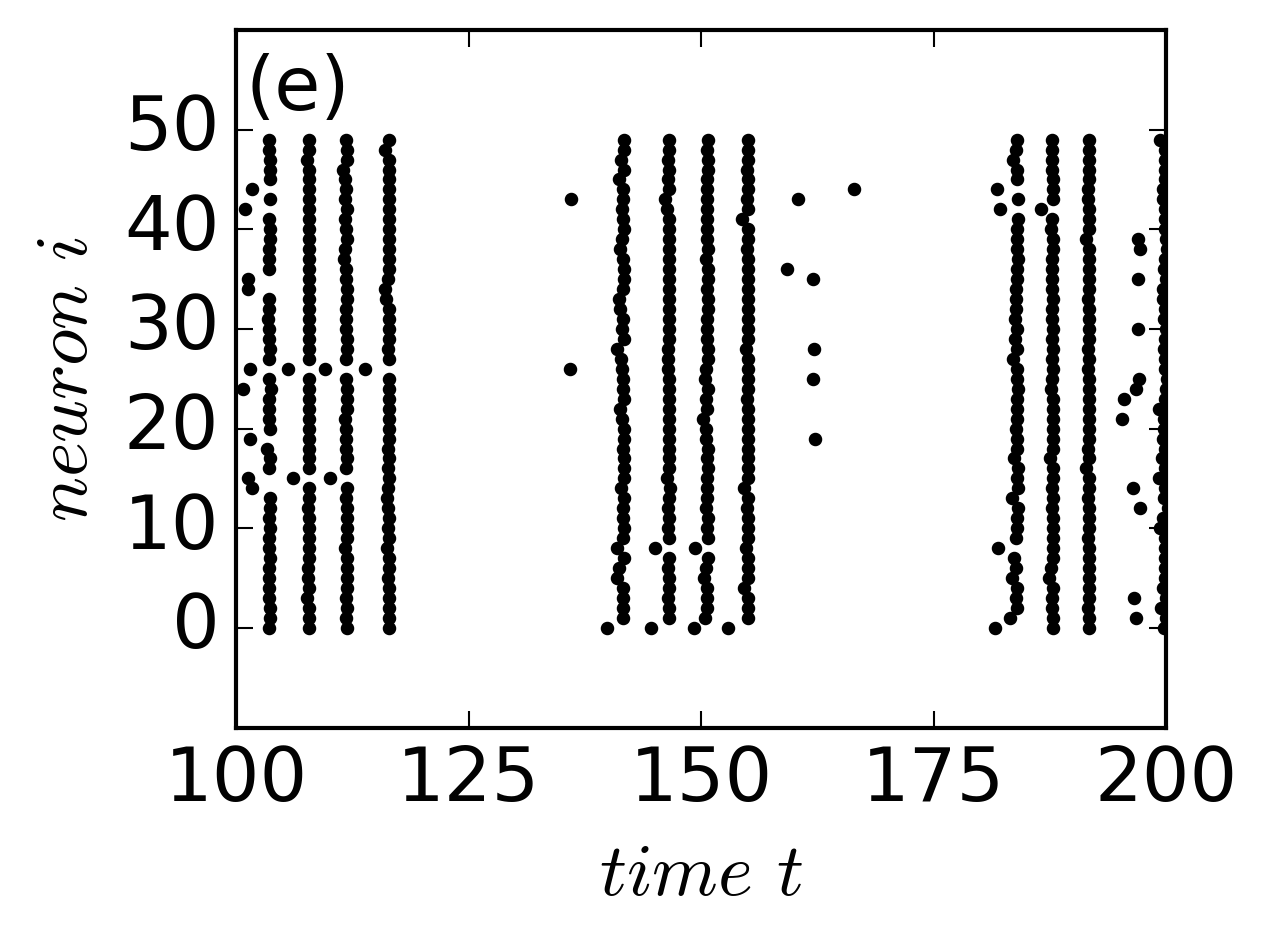}
\caption{Raster plots displaying the spiking activity of the group of 50 neurons for the same parameters as in Fig.~\ref{Fig0} }
\label{Fig2}
\end{figure}

The neuronal ensemble displays different dynamical regimes, depending on the coupling strengh, the signal amplitud and period, the noise strength, and the coupling topology. Figures~\ref{Fig0} and \ref{Fig2} display several examples of the dynamics of a group of 50 neurons under different conditions: Fig.~\ref{Fig0} shows the activity of an individual neuron (the voltage-like variable of neuron 1), while Fig.~\ref{Fig2} displays the raster plot of the ensemble. In panels \ref{Fig0}(a) and \ref{Fig2}(a)  the neurons are uncoupled and no signal is applied, therefore, random spiking activity occurs due to the noise. In panels \ref{Fig0}(b) and \ref{Fig2}(b) the neurons are mutually coupled, still no signal is applied. Now we see synchronized spiking activity superposed with random spikes. When the periodic signal is applied, we see in panels \ref{Fig0}(c)-(f) and \ref{Fig2}(c)-(f) that the neurons either fire regular and synchronized spikes, or there is more irregular firing, depending on the period of the signal.

In the following we analyze the influence of the different parameters. To stress the role of the number of neurons, we compare the results obtained for 50 neurons with those obtained for only two coupled neurons. 

\begin{figure}[!ht]
\begin{subfigure}{.5\textwidth}
  \centering
\includegraphics[width=.95\linewidth]{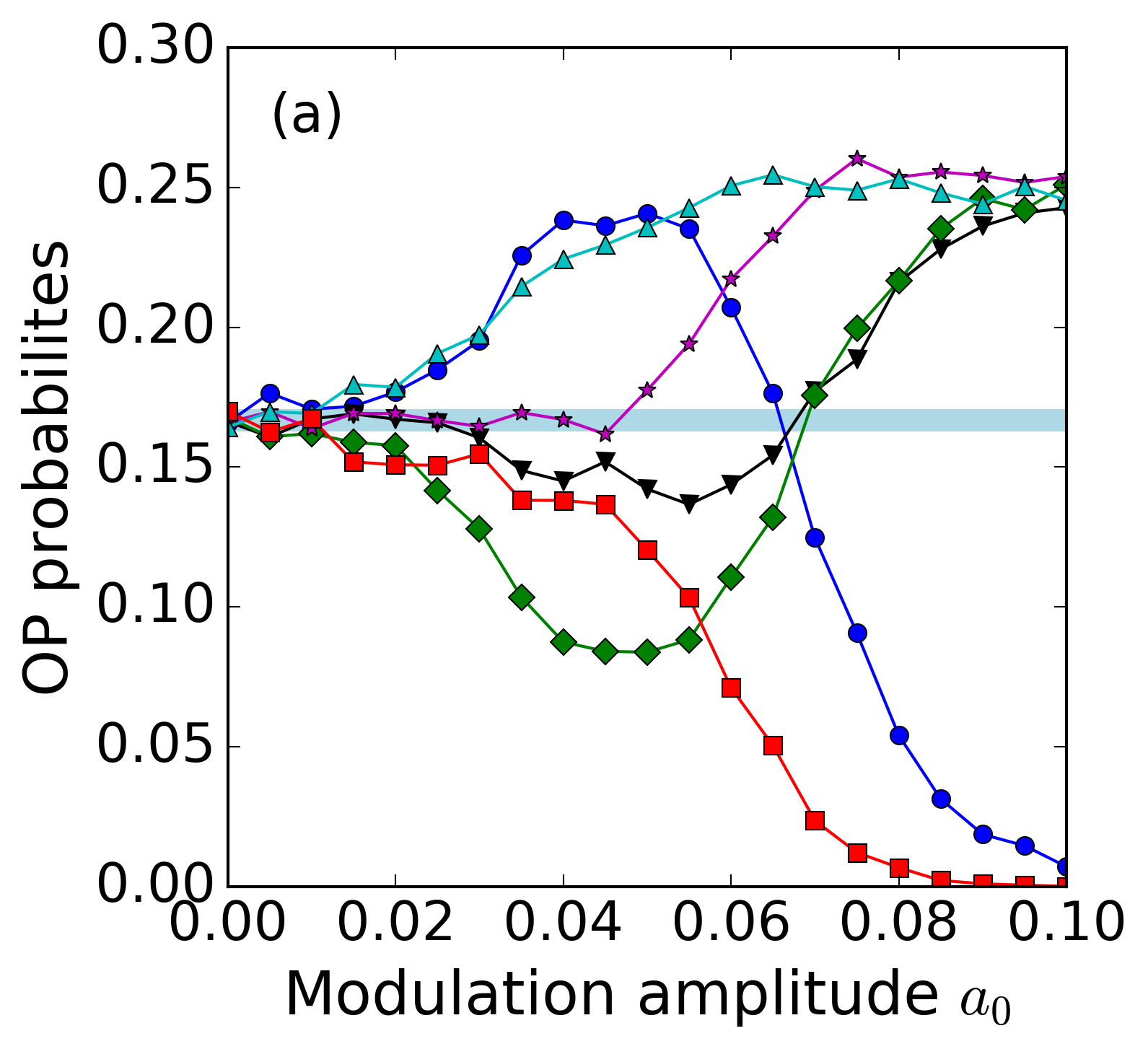}
\end{subfigure}%
\begin{subfigure}{.5\textwidth}
  \centering
 \includegraphics[width=.95\linewidth]{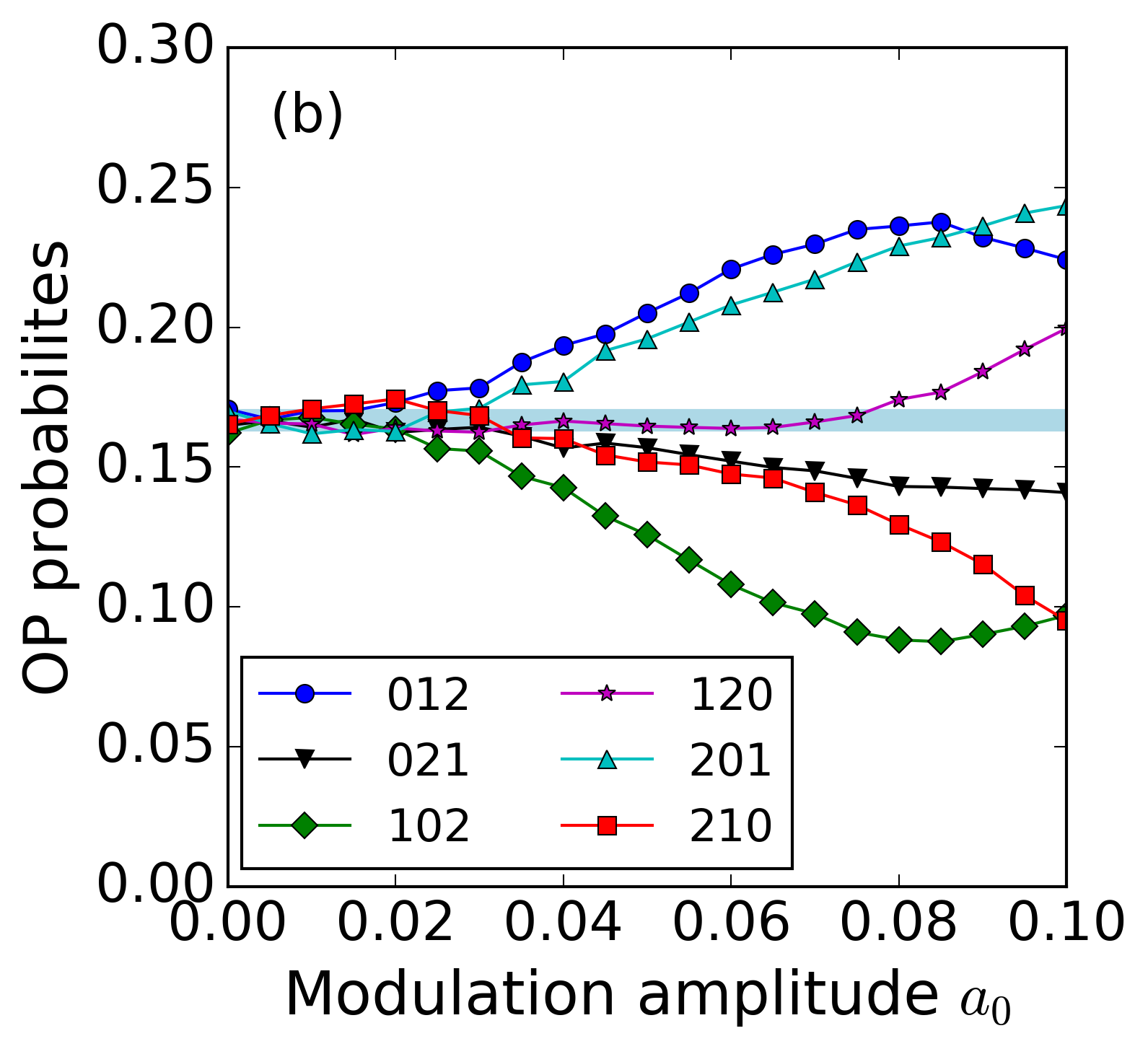}
\end{subfigure}
\caption{Probabilities of the ordinal patterns as a function of the signal amplitude, $a_0$, for (a) an ensemble of 50 neurons, all-to-all coupled, and for (b) two mutually coupled neurons. The parameters are: $T = 10$, $D = 2.5\cdot 10^{-6}$ and $\sigma = 0.05$.}
\label{Fig1}
\end{figure}

We begin by characterizing the role of the signal amplitude, presented in Fig. \ref{Fig1} that displays probabilities of the six ordinal patterns as a function of $a_0$ for $N = 50$ [Fig. \ref{Fig1} (a)] and for $N = 2$ [Fig. \ref{Fig1} (b)].  Here $a_0$ is kept within the range of values for which, in the absence of noise, the neurons do not fire spikes. 
We note that, if the signal amplitude is small enough, as expected, the ordinal probabilities are within the blue region that indicates values that are consistent with equal probabilities, with 99.74$\%$ confidence level (this region is calculated as explained in Sec.~\ref{Methods}). 

\begin{figure}[!ht]
\begin{subfigure}{.5\textwidth}
  \centering
 \includegraphics[width=.95\linewidth]{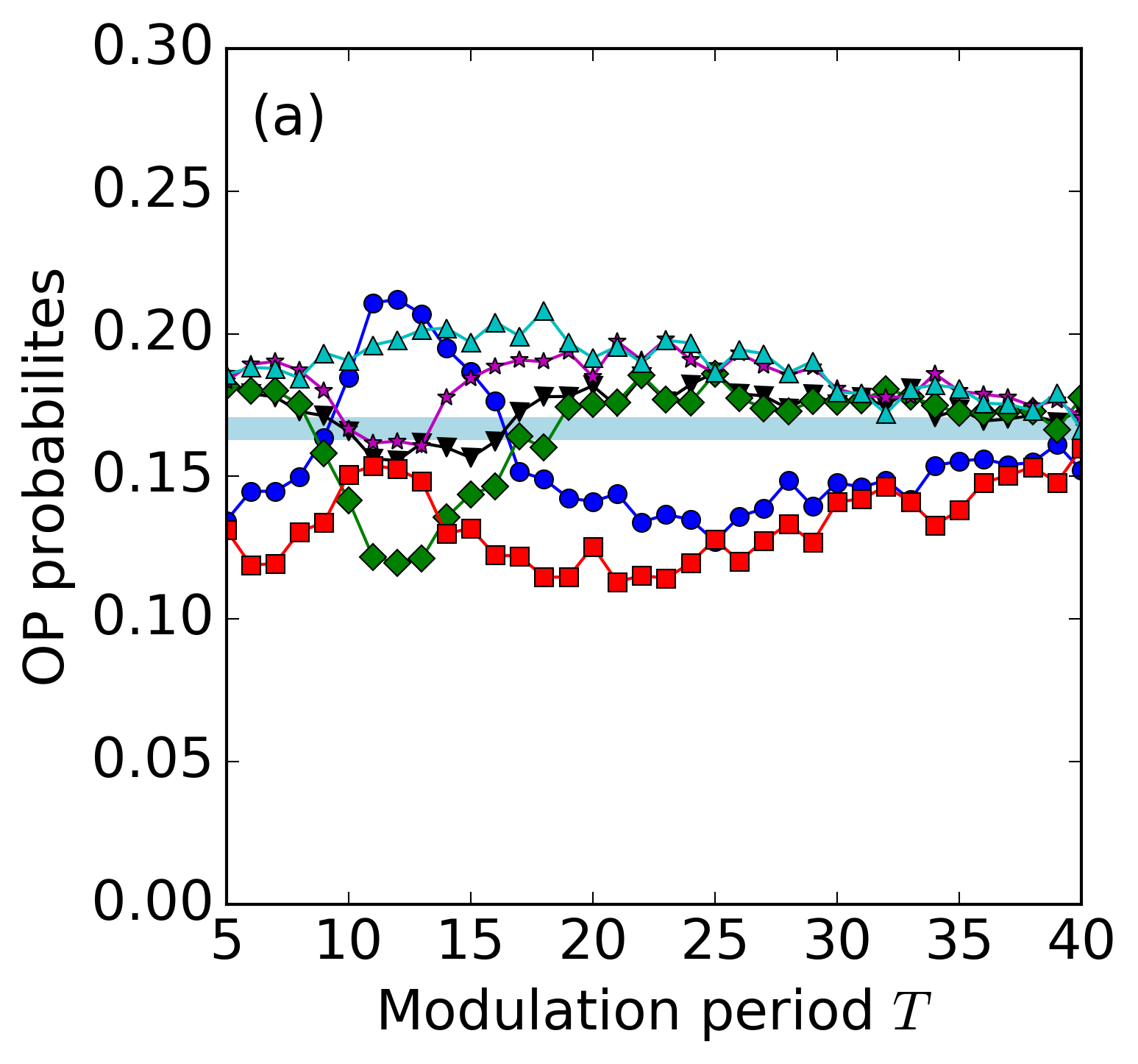}
\end{subfigure}%
\begin{subfigure}{.5\textwidth}
  \centering
 \includegraphics[width=.95\linewidth]{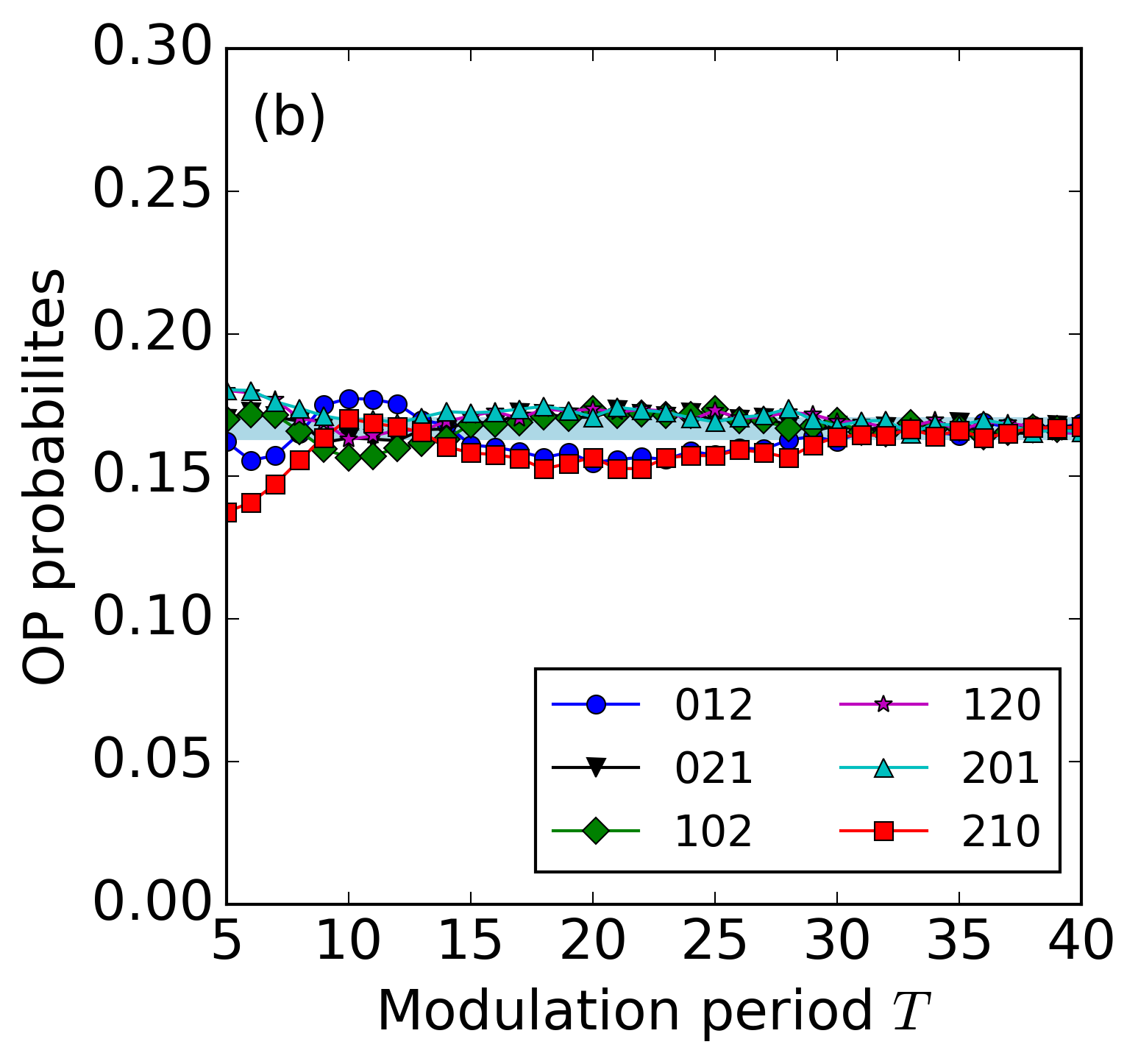}
\end{subfigure}
\begin{subfigure}{.5\textwidth}
  \centering
 \includegraphics[width=.95\linewidth]{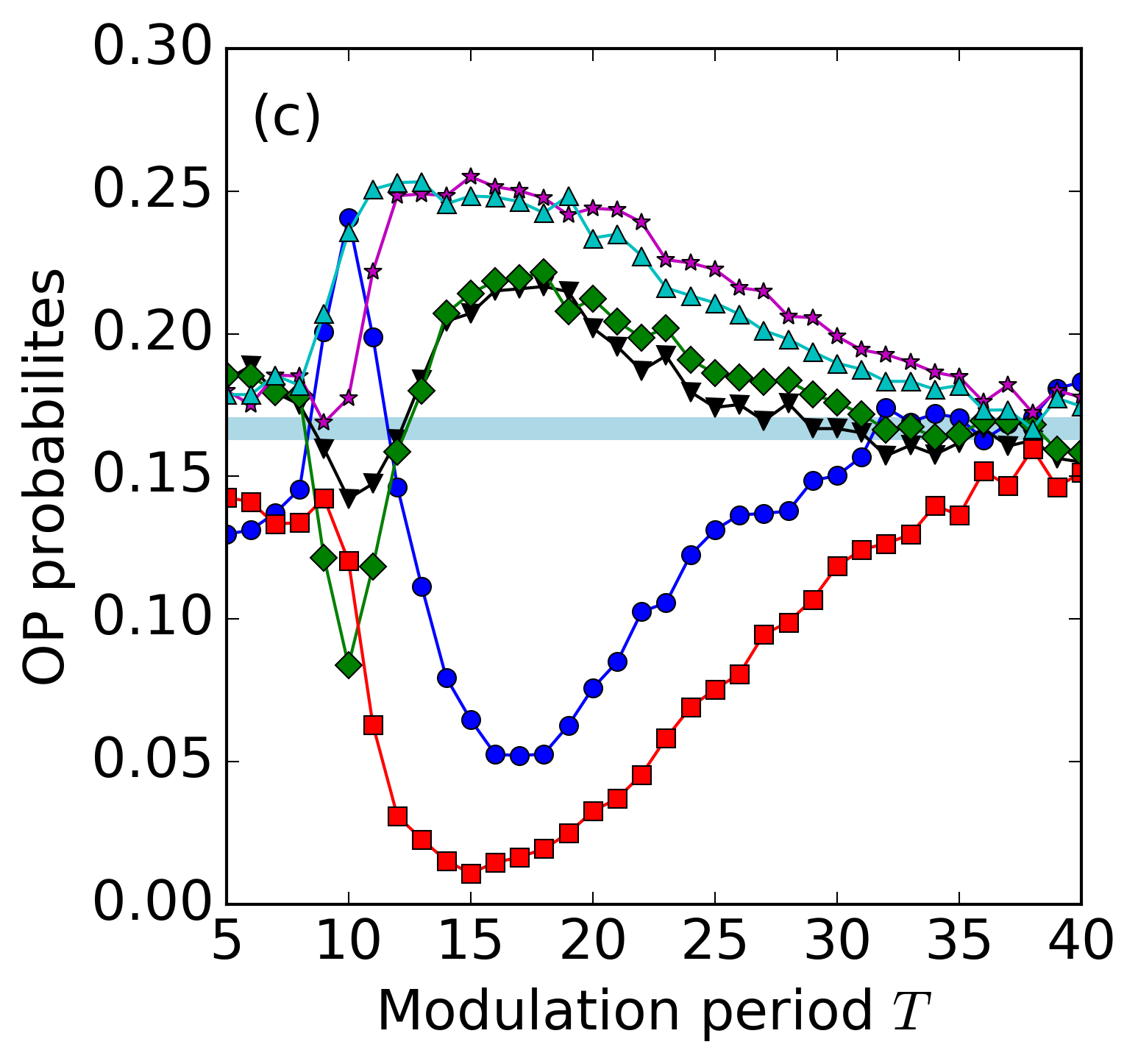}
\end{subfigure}
\begin{subfigure}{.5\textwidth}
  \centering
  \includegraphics[width=.95\linewidth]{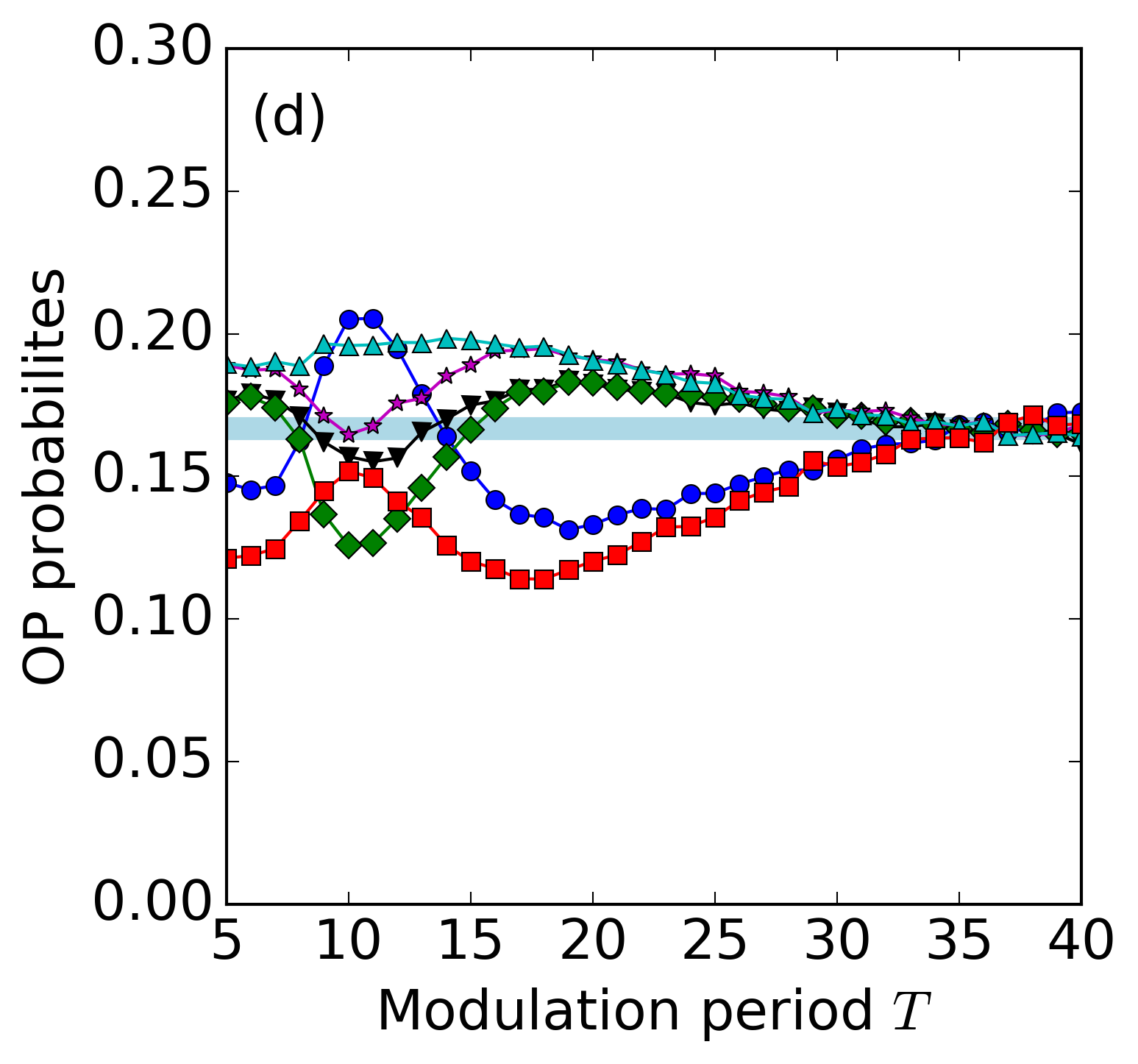}
\end{subfigure}
\begin{subfigure}{.5\textwidth}
  \centering
  \includegraphics[width=.95\linewidth]{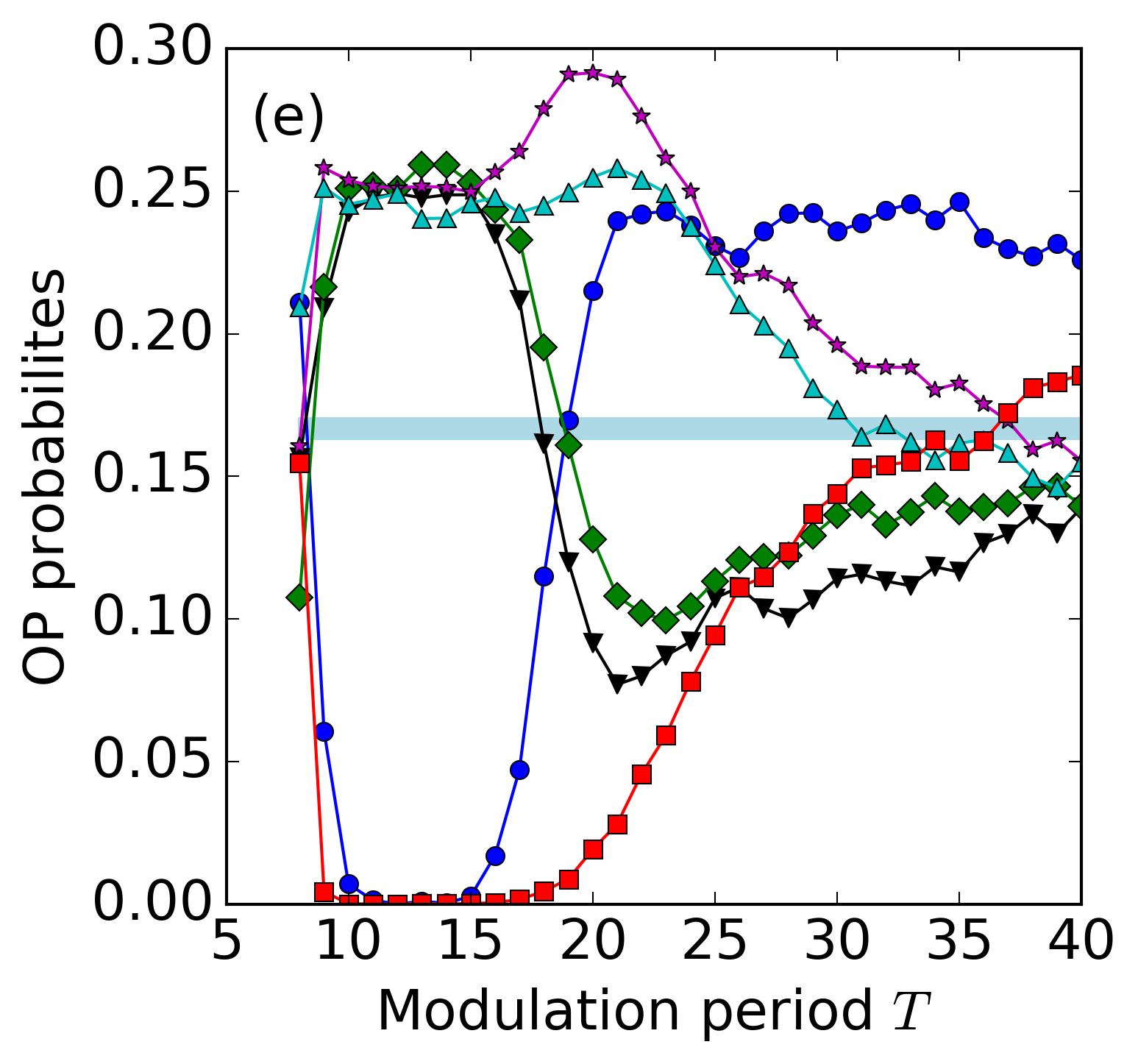}
\end{subfigure}
\begin{subfigure}{.5\textwidth}
  \centering
  \includegraphics[width=.95\linewidth]{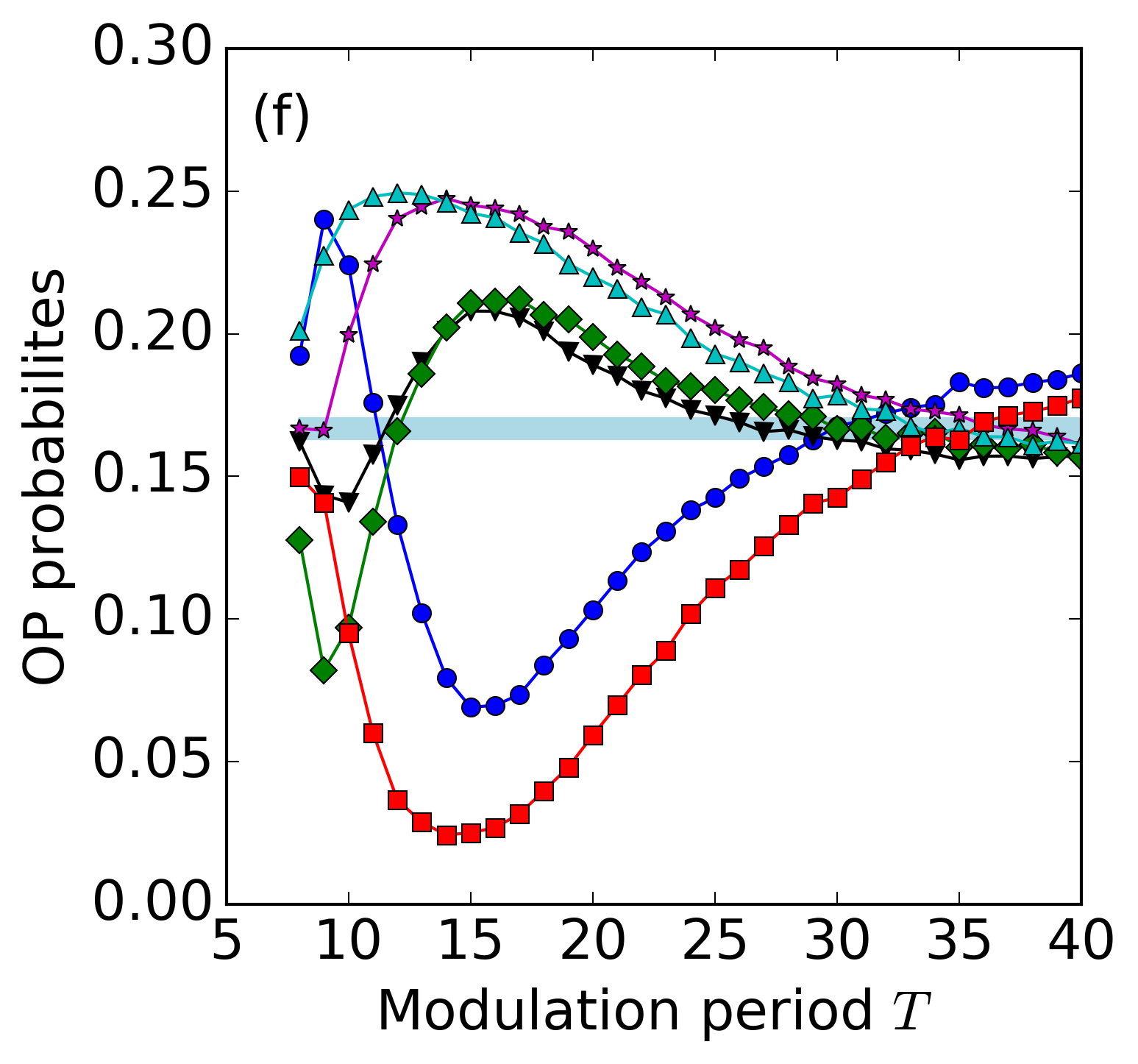}
\end{subfigure}
\caption{Probabilities of the ordinal patterns as a function of the signal period, $T$, for (a,b) $a_0 = 0.025$, (c,d) $a_0 = 0.05$ and (e,f) $a_0 = 0.1$ with $N = 50$ (a,c,e) and $N = 2$ (b,d,f). In panels (e) and (f) we consider $T \ge 8$ because for $T < 8$ the signal by itself triggers spikes. Other parameters are: $D = 2.5\cdot 10^{-6}$ and $\sigma = 0.05$.}
\label{Fig3}
\end{figure}

As the signal amplitude increases we note that, while for the two coupled neurons, ordinal probabilities gradually increase (or decrease), for the ensemble of 50 neurons their variation is more pronounced. Interestingly, the same codification (i.e., same ordinal patterns probabilities) is obtained for $N=2$ and larger $a_0$. This is shown in Fig. \ref{Fig3}, where we analyze the effect of the signal period ($T$ is kept within the range of values for which, in the absence of noise, the neurons only display sub-threshold oscillations): comparing Figs. \ref{Fig3} (a) and \ref{Fig3} (d), or Figs. \ref{Fig3} (c) and \ref{Fig3} (f), we see that for two neurons and larger signal amplitudes we find a very similar set of ordinal probabilities as for 50 neurons and lower $a_0$. 
We see that the variation of the ordinal probabilities with the period is very similar for $a_0=0.025, N=50$ and $a_0=0.05, N=2$ [in Figs. \ref{Fig3} (a) and \ref{Fig3} (d), respectively] and for $a_0=0.05, N=50$ and $a_0=0.1, N=2$ [Figs. \ref{Fig3} (c) and\ref{Fig3} (f), respectively]. Therefore, these results suggest that 50 neurons encode a weak signal in a very similar way as 2 neurons encode a stronger signal.

Regarding how the encoding of the signal depends on its period, in Fig. \ref{Fig3} we verify that the probabilities of the patterns expressed in the spike sequences depend on the period of the signal (consistent with the observations in \cite{REI16,MAS18}). Comparing the left and right columns of Fig. \ref{Fig3}, we note that neuronal coupling is beneficial for signal encoding because for $N = 50$ (left column) the ordinal probabilities take higher or lower values, and the resonances with the period become more pronounced, as compared to $N = 2$ (right column). 

Interestingly, for $N=50$ and $a_0 = 0.1$ the probabilities are nearly constant in the interval $10 \le T \le 15$ and patterns 012 and 210 have very low or zero probability. The corresponding neuronal activity for $T=10$ was displayed in Figs.~\ref{Fig0}(c) and \ref{Fig2}(c). We see an alternation of long and short intervals between spikes, while three consecutive increasing or decreasing intervals do not occur (which would be represented by patterns 012 and 210 respectively).

\begin{figure}[!ht]
\begin{subfigure}{.5\textwidth}
  \centering
\includegraphics[width=.95\linewidth]{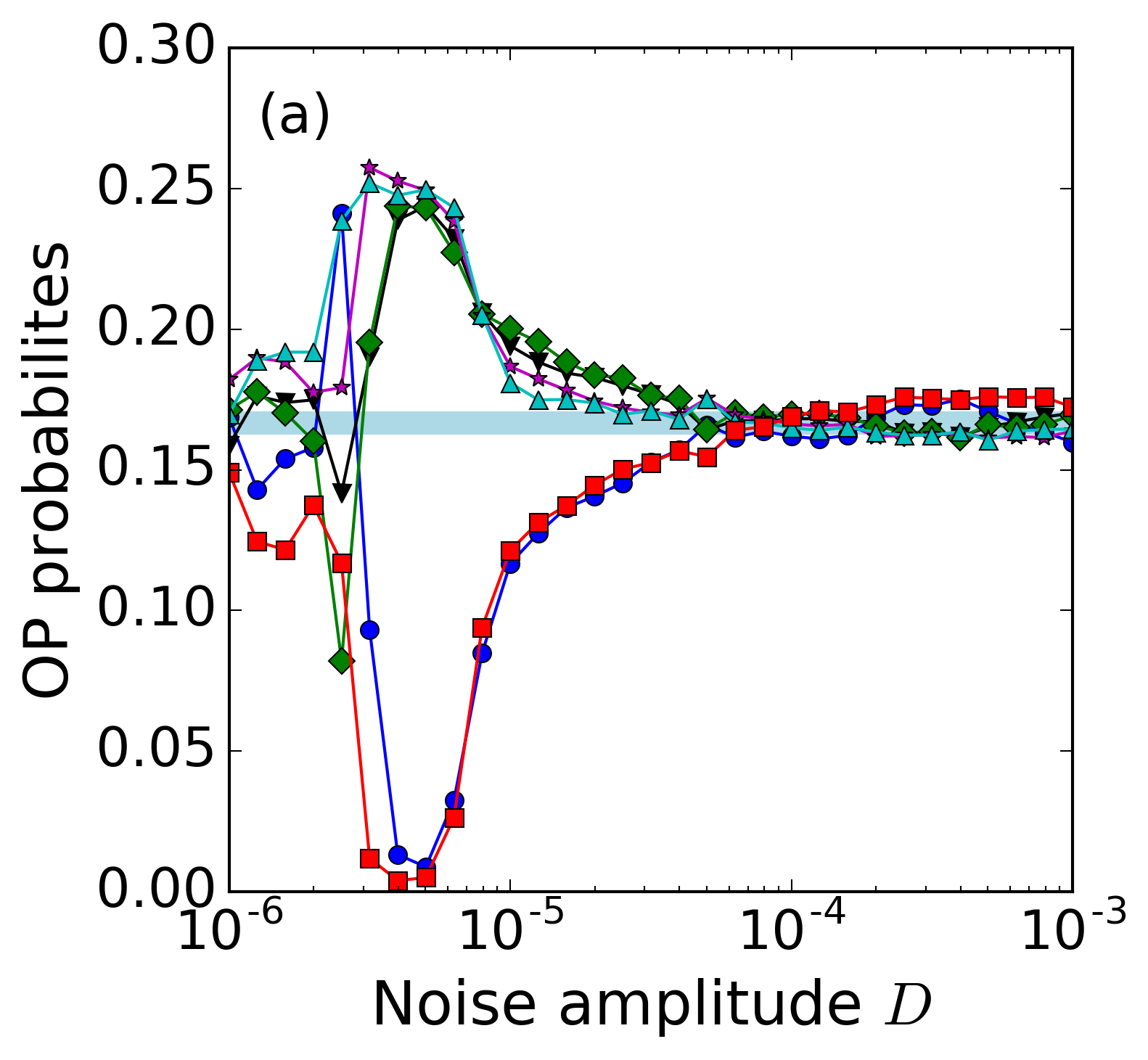}
\end{subfigure}%
\begin{subfigure}{.5\textwidth}
  \centering
\includegraphics[width=.95\linewidth]{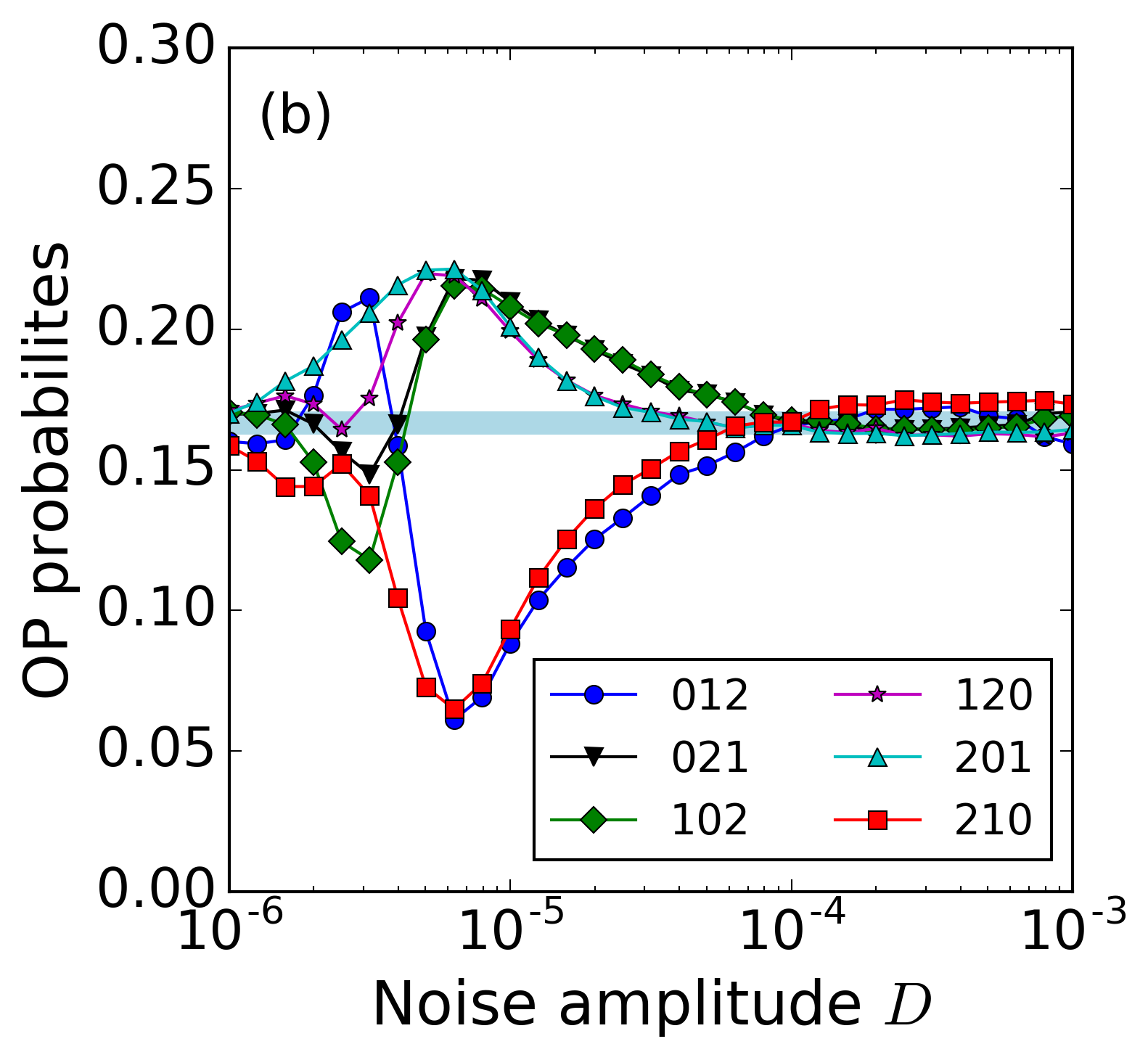}
\end{subfigure}
\centering
\caption{Probabilities of the ordinal patterns as a function of the noise strength, $D$, for (a) 50 neurons and for (b) two neurons. Other parameters are: $a_0 = 0.05$, $T = 10$,  and $\sigma = 0.05$.}
\label{Fig6}
\end{figure}

In Ref.~\cite{REI16} it was shown that the ordinal patterns displayed a noise-induce resonance, as 012 and 210 reached minimum values when the noise intensity was such that the mean ISI, $\langle ISI \rangle$, was approximately equal to half the signal period. In Ref.~\cite{MAS18} it was demonstrated that this encoding mechanism persisted when the neuron was coupled to a second neuron that did not perceive the signal. Here, we show in Fig. \ref{Fig6}(a) that the mechanism is robust and the resonance is more pronounced when the signal is perceived by a group of 50 neurons: ordinal patterns 012 and 210 are not expressed (have zero probability) when $D = 5\cdot 10 ^{-6}$, and for this noise strength, $\langle ISI \rangle  =T/2$. For comparison Fig. \ref{Fig6}(b) shows the ordinal probabilities as a function of $D$ for $N = 2$. Ordinal patterns 012 and 210 are minimum for almost the same noise strength ($D = 8\cdot 10 ^{-6}$) which gives $\langle ISI \rangle  = T/2$. Yet, the minimum is less pronounced, as compared to the group of 50 neurons.

So far we have seen that for 50 neurons the encoding of the signal is, in general, improved in comparison with that of only two neurons. Yet, how is the variation of the ordinal probabilities with the network size? Next, we fix the period and the amplitude of the signal and we characterize the influence of i) the number of neurons, $N$, when they are all-to-all coupled; ii) the number of links (from zero links to all-to-all coupling, randomly adding links) and iii) the strength of the coupling, $\sigma$, in the all-to-all configuration, from 0 (uncoupled neurons) to the same coupling strength considered in steps i) and ii). We keep the coupling level low enough such that, without signal and noise, there are no spikes. We note that the starting and final points in the three steps are the same: from the uncoupled neurons to 50 all-to-all coupled neurons. 

\begin{figure}[!ht]
\begin{subfigure}{.5\textwidth}
  \centering
 \includegraphics[width=.95\linewidth]{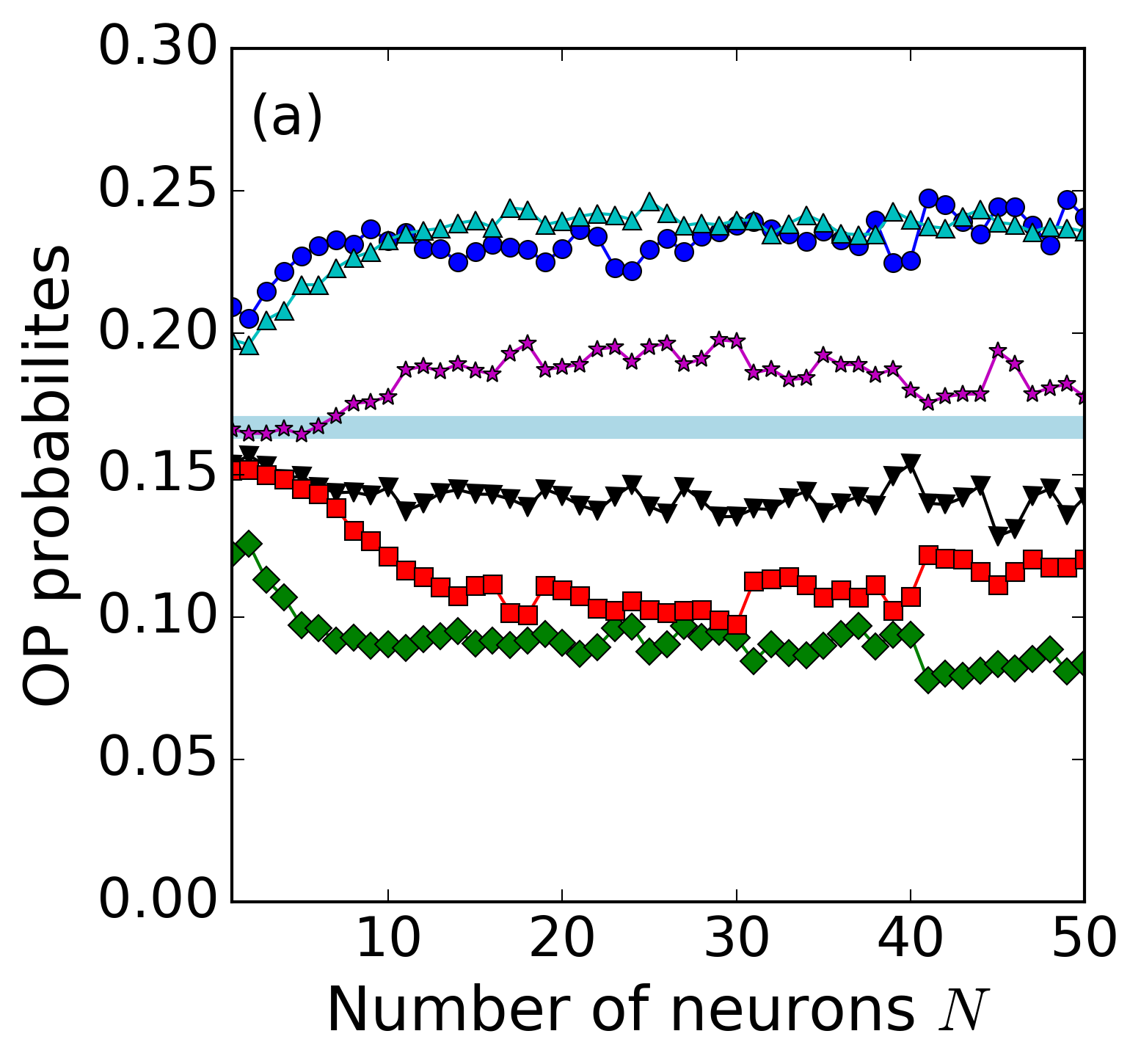}
\end{subfigure}%
\begin{subfigure}{.5\textwidth}
  \centering
 \includegraphics[width=.95\linewidth]{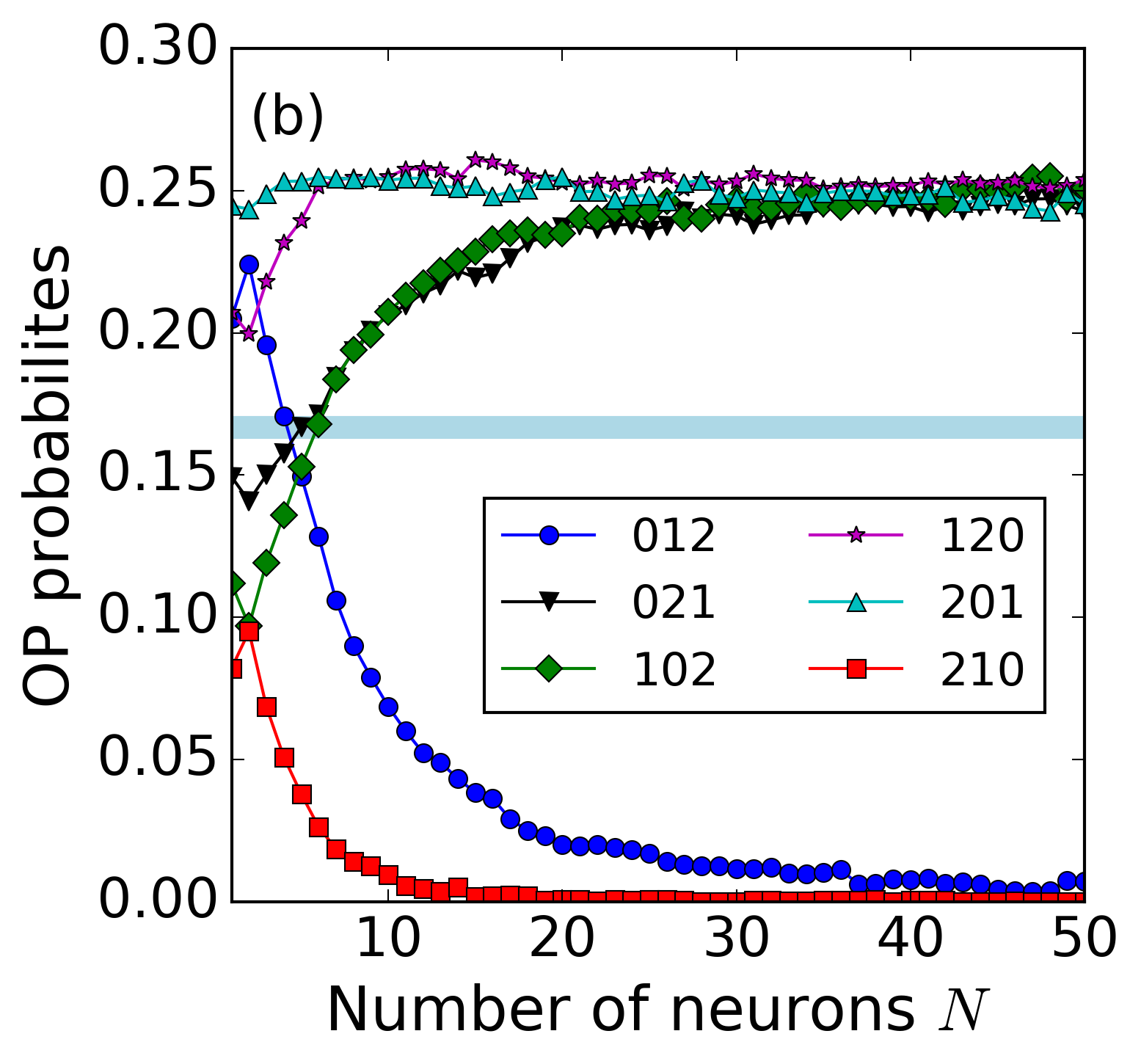}
\end{subfigure}
\begin{subfigure}{.5\textwidth}
  \centering
  \includegraphics[width=.95\linewidth]{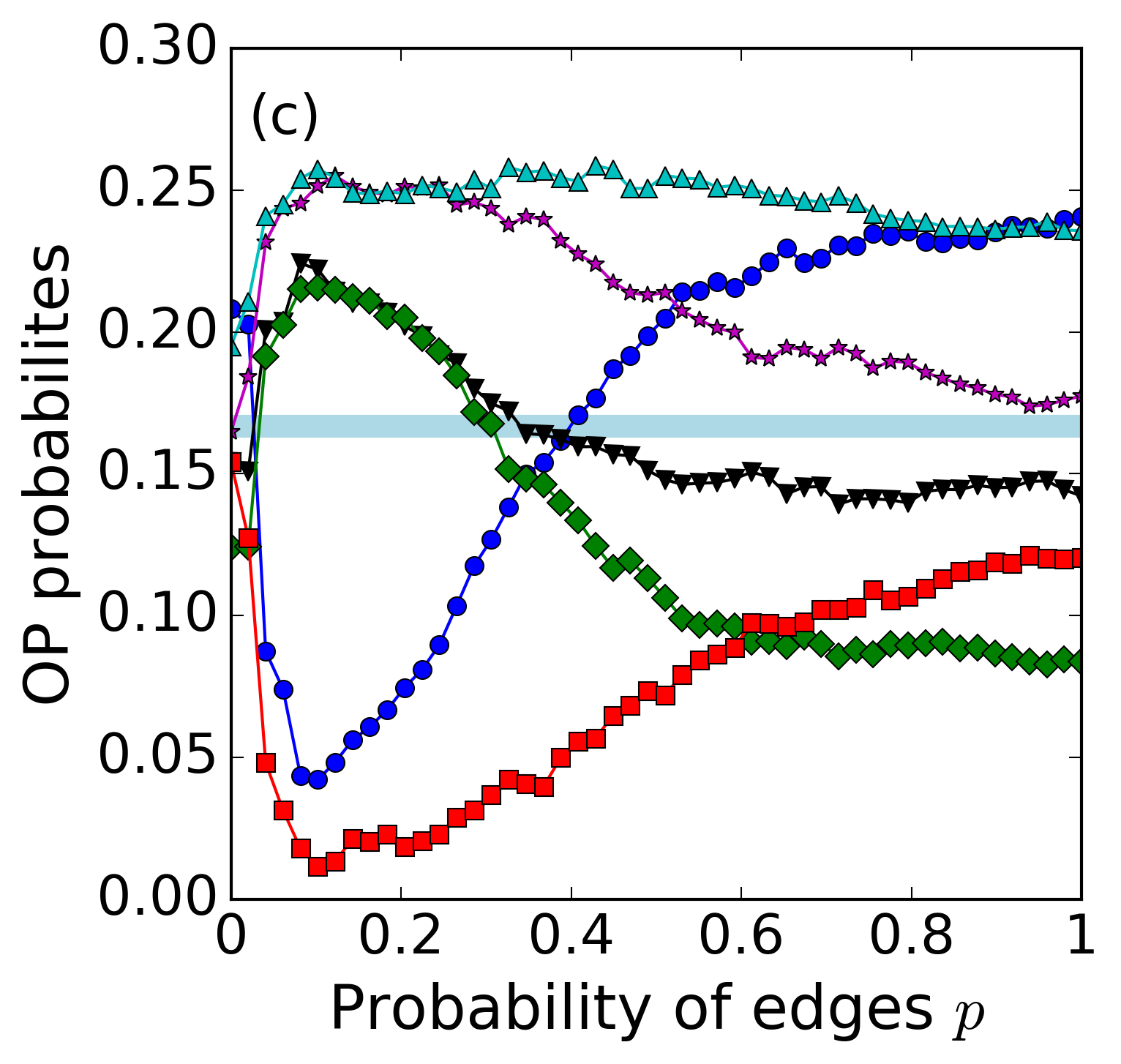}
\end{subfigure}
\begin{subfigure}{.5\textwidth}
  \centering
  \includegraphics[width=.95\linewidth]{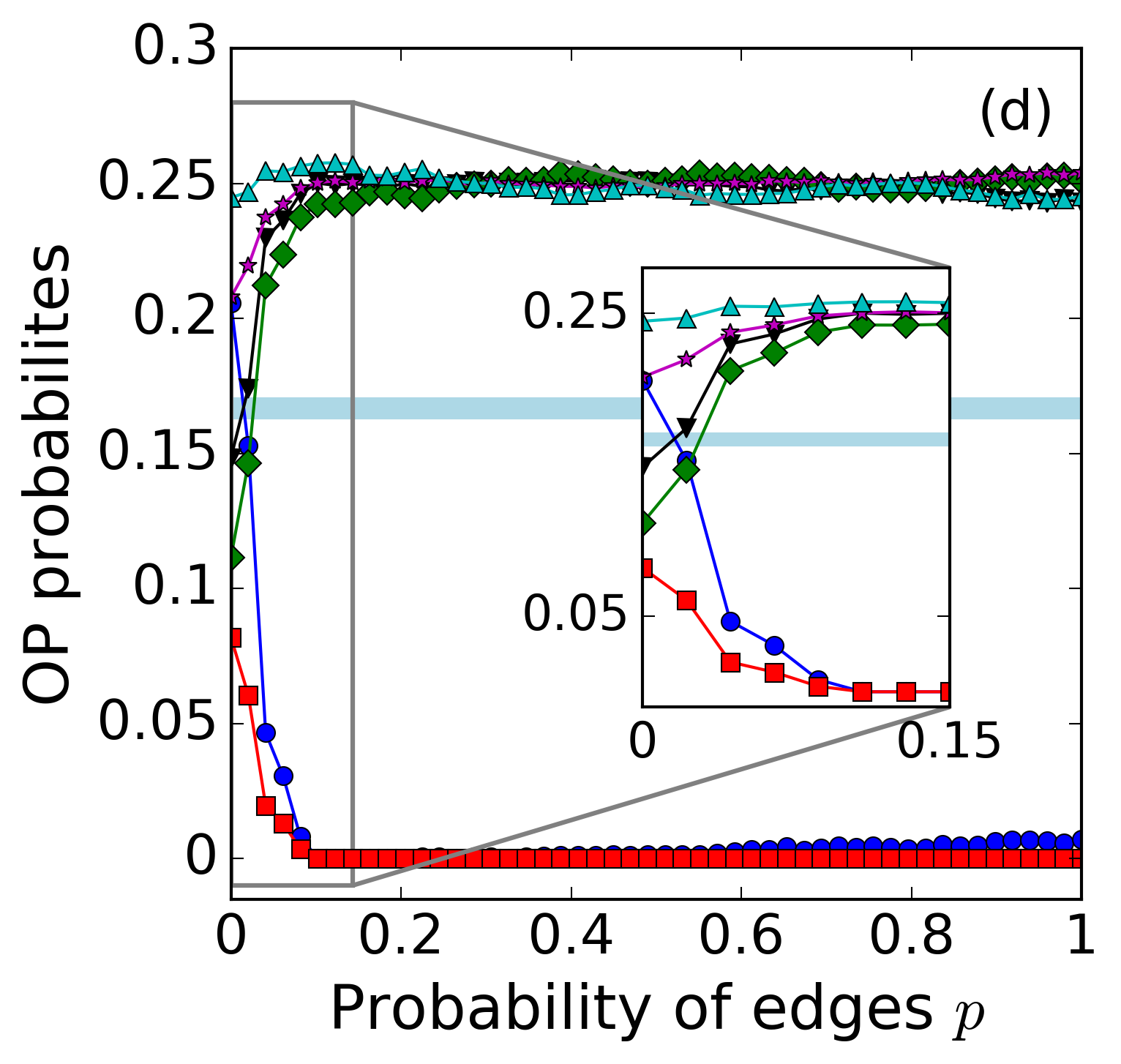}
\end{subfigure}
\begin{subfigure}{.5\textwidth}
  \centering
  \includegraphics[width=.95\linewidth]{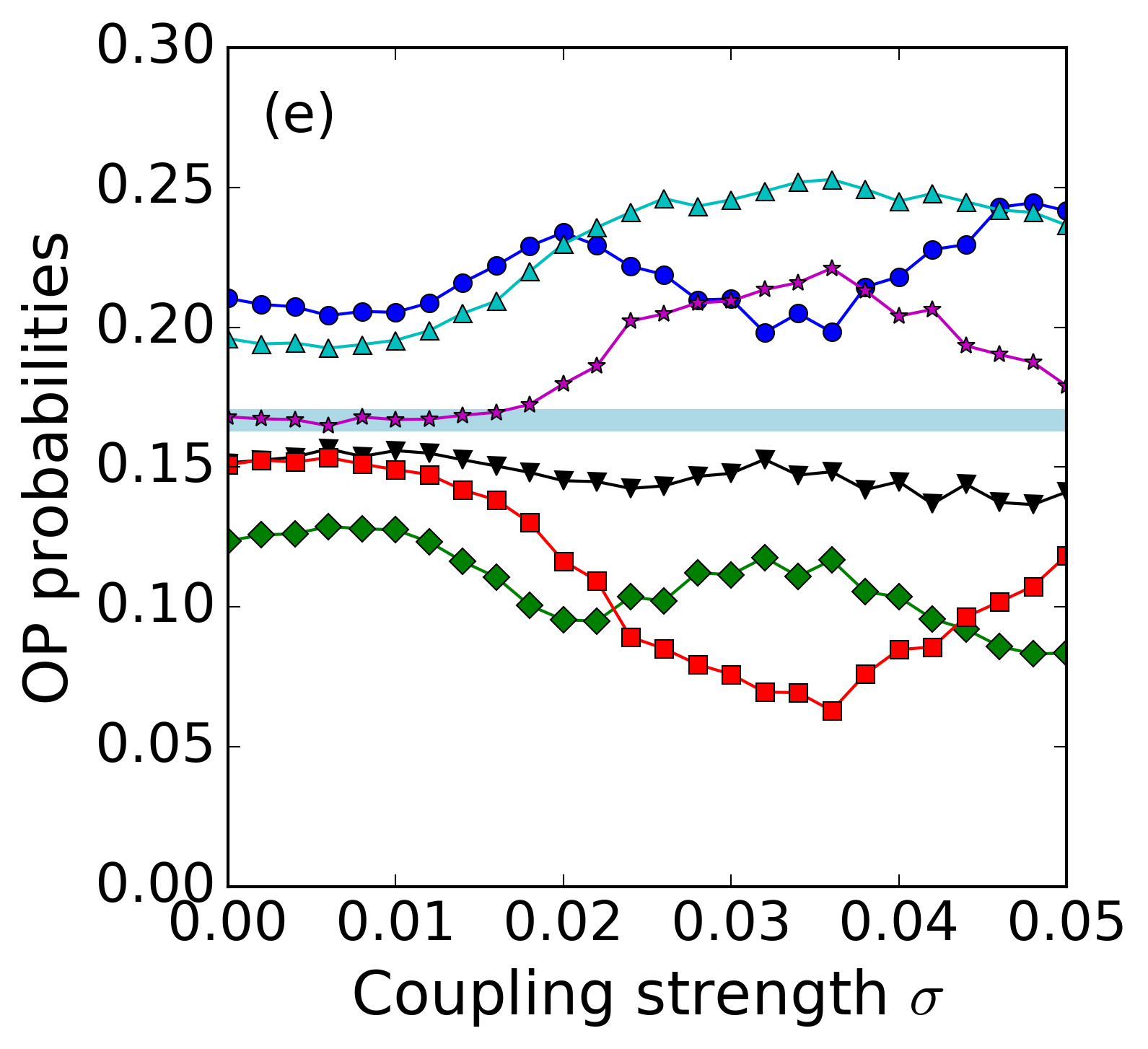}
\end{subfigure}
\begin{subfigure}{.5\textwidth}
  \centering
  \includegraphics[width=.95\linewidth]{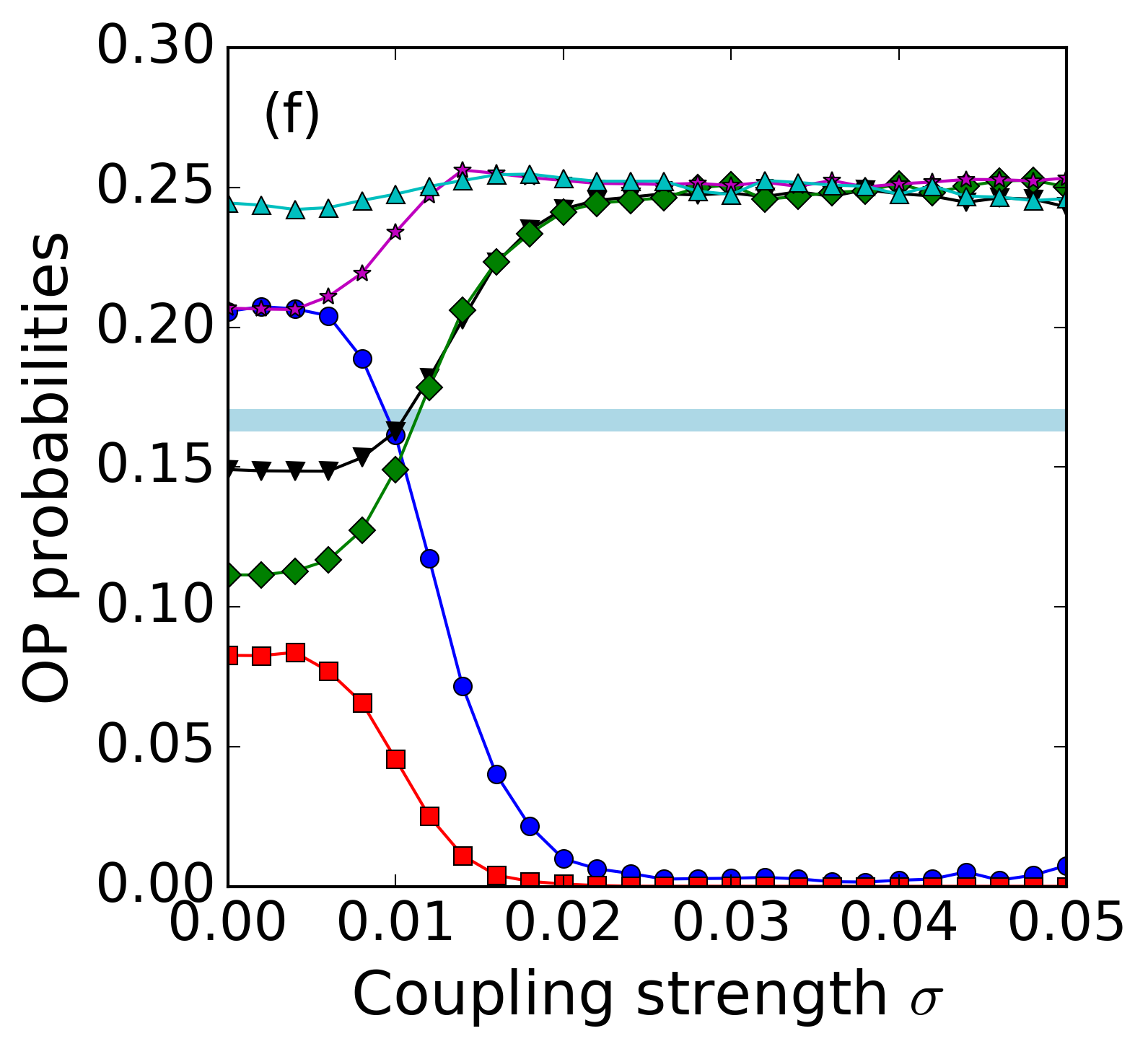}
\end{subfigure}
\caption{Probabilities of the ordinal patterns as a function of the number of neurons, $N$ (a, b), of the percentage of links (c, d), and of the coupling strength (e, f) for $a_0 = 0.05$ and $a_0 = 0.1$, respectively. In panels (a, b, e, f) the neurons are all-to-all coupled, in panels (c, d) the coupling topology is random (starting from uncoupled neurons, links are randomly added until the neurons are all-to-all coupled). In panels (c, d, e, f) $N=50$, in all the panels: $D = 2.5\cdot 10^{-6}$ and $T = 10$.}
\label{Fig4}
\end{figure}

Figure \ref{Fig4} presents the results: panels (a, b)  display the ordinal probabilities as a function of $N$; (c, d) as a function of the percentage of total links; and (e, f) as a function of the coupling strength. To investigate if these parameters can play different roles for weak or strong signals, we consider two signal amplitudes: $a_0 = 0.05$ in panels (a, c, e) and $a_0 = 0.1$ in panels (b, d, f).

In Fig. \ref{Fig4}(a) we note that for $a_0 = 0.05$ the probabilities gradually vary, increasing or decreasing, as $N$ increases up to $N = 10$. With further increase of $N$ they remain nearly constant. The signal is encoded (the probabilities are not in the blue region) but, at least for these parameters, the encoding only slightly improves when increasing $N$. In contrast, for $a_0 = 0.1$ [Fig. \ref{Fig4}(b)] we observe that the encoding is significantly improved, compared to Fig. \ref{Fig4}(a), as the probabilities of the ordinal patterns 012 and 210 gradually decrease to zero. An interesting observation is that above a certain number of neurons (which depends on the parameters) the probabilities saturate and remain stable with further increase of $N$.
 
\begin{figure}[!ht]
\begin{subfigure}{.5\textwidth}
  \centering
 \includegraphics[width=.95\linewidth]{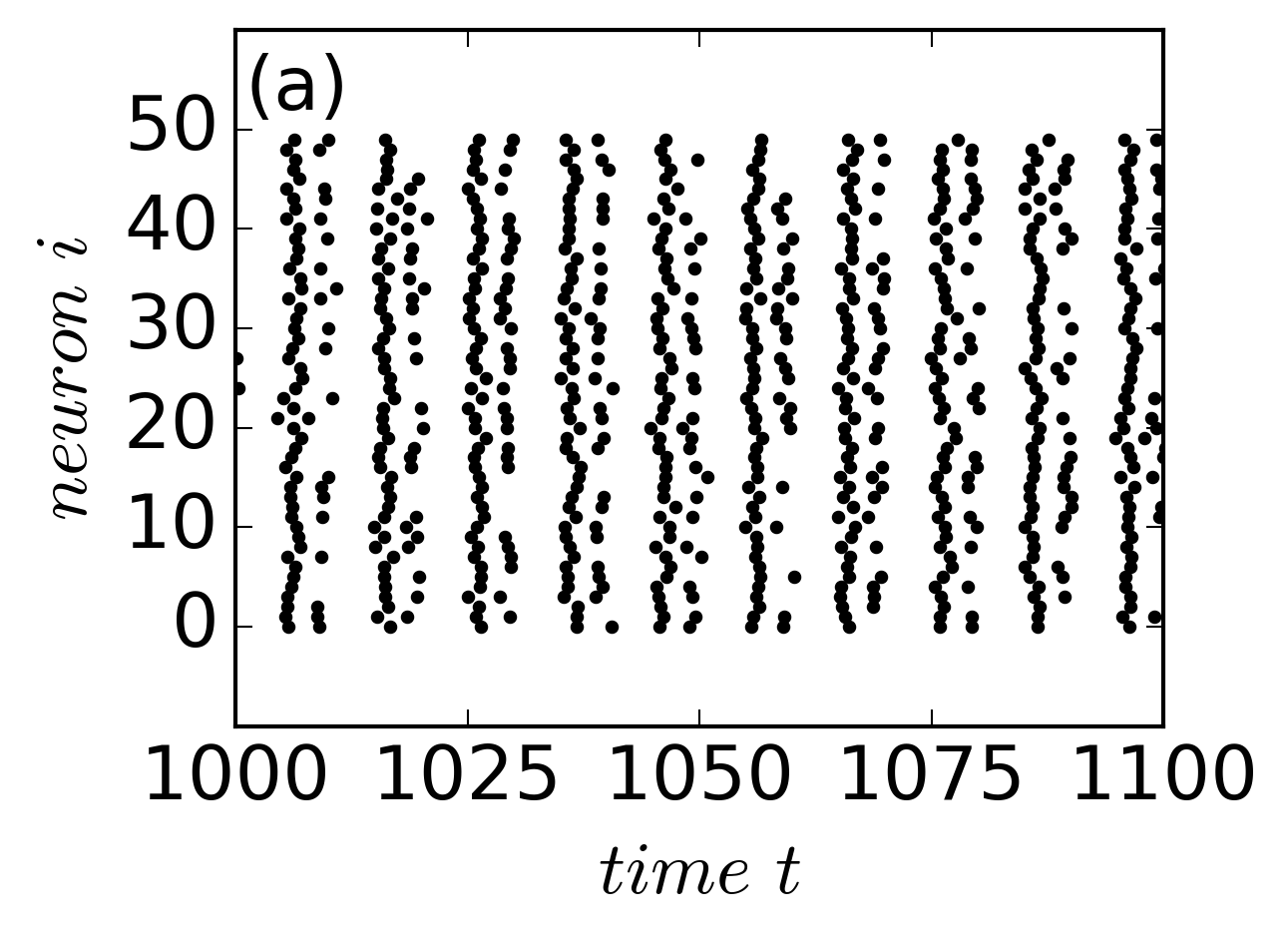}
\end{subfigure}%
\begin{subfigure}{.5\textwidth}
  \centering
  \includegraphics[width=.95\linewidth]{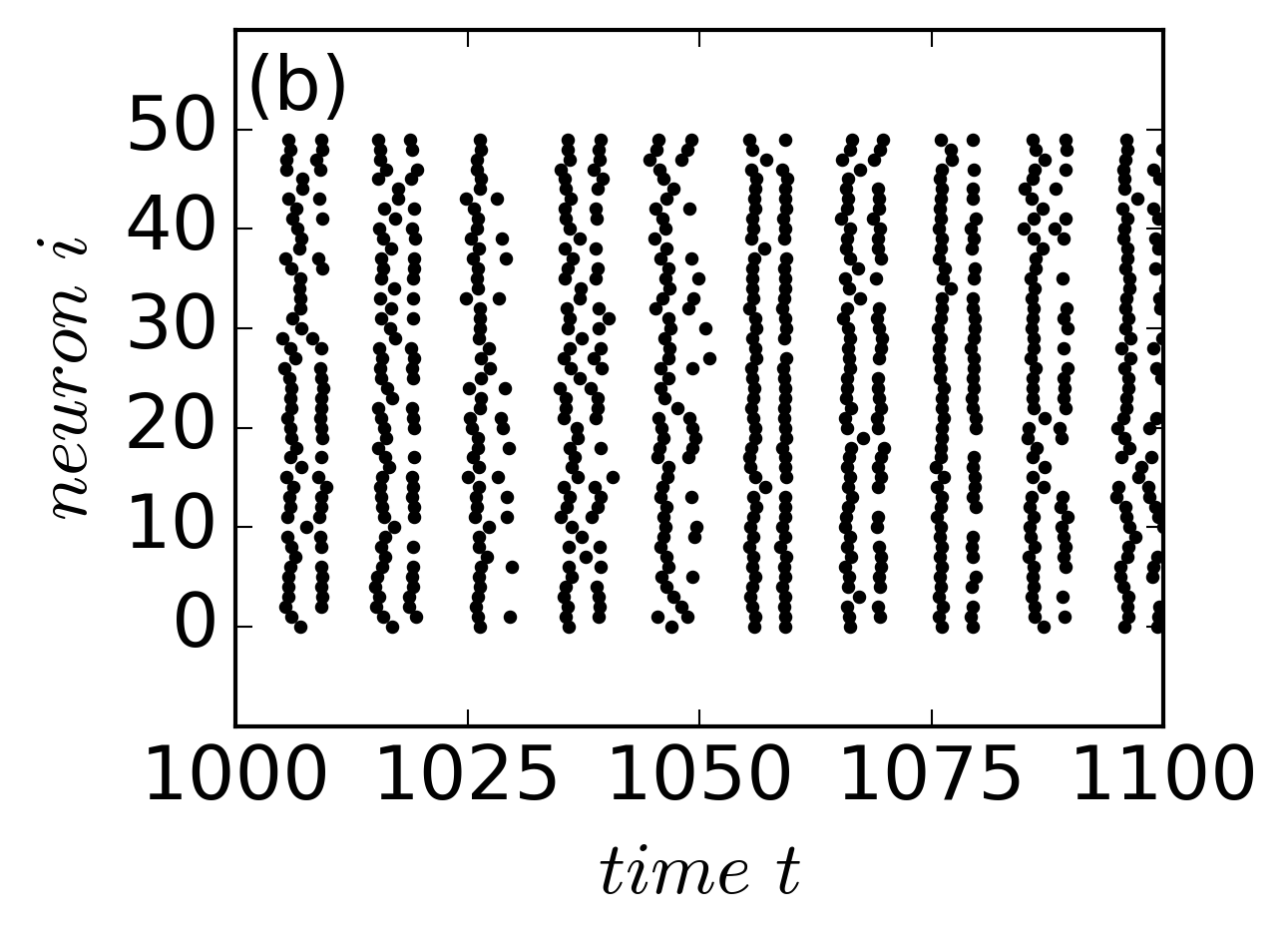}
\end{subfigure}
\begin{subfigure}{.5\textwidth}
  \centering
  \includegraphics[width=.95\linewidth]{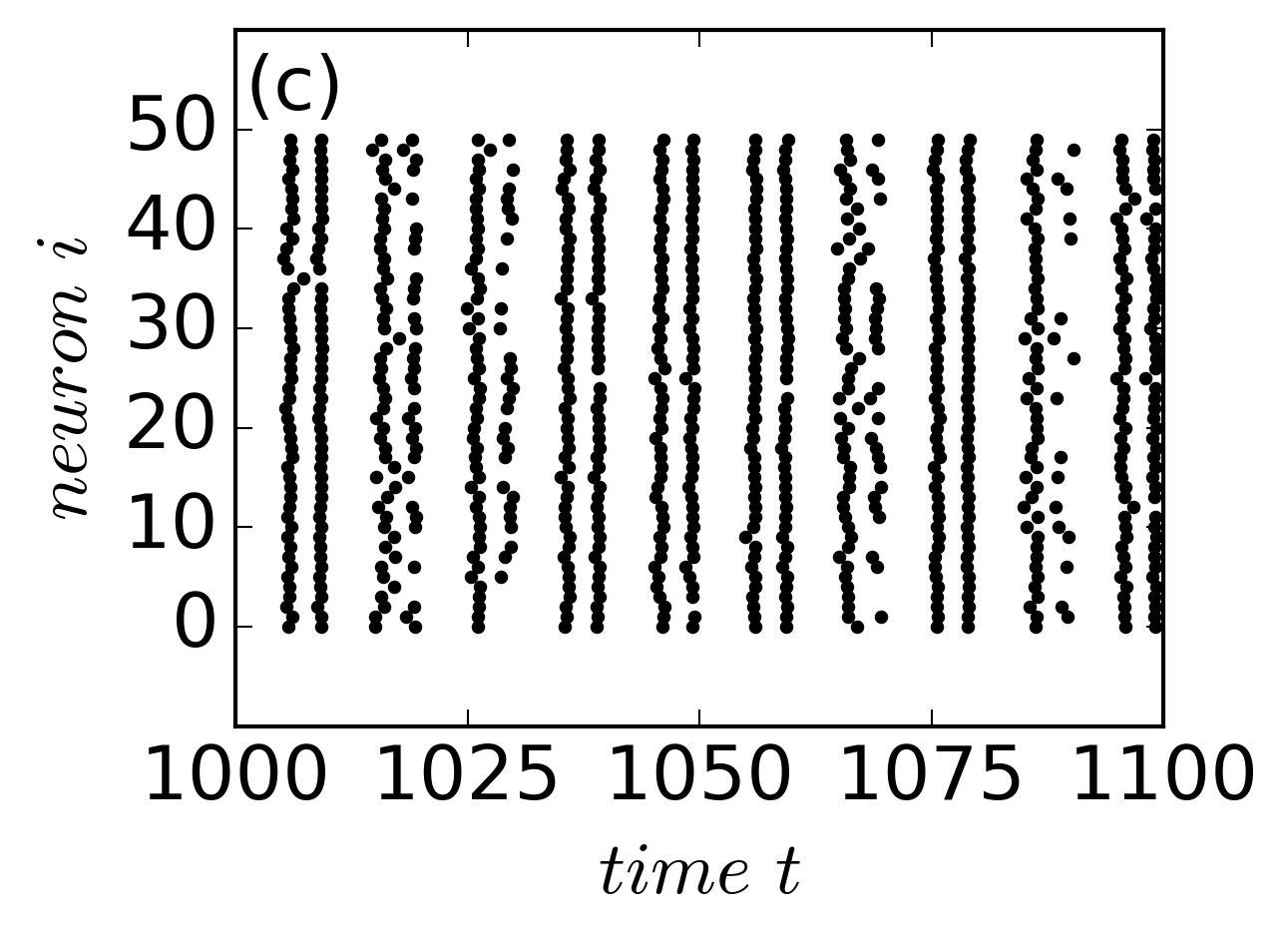}
\end{subfigure}
\begin{subfigure}{.5\textwidth}
  \centering
  \includegraphics[width=.95\linewidth]{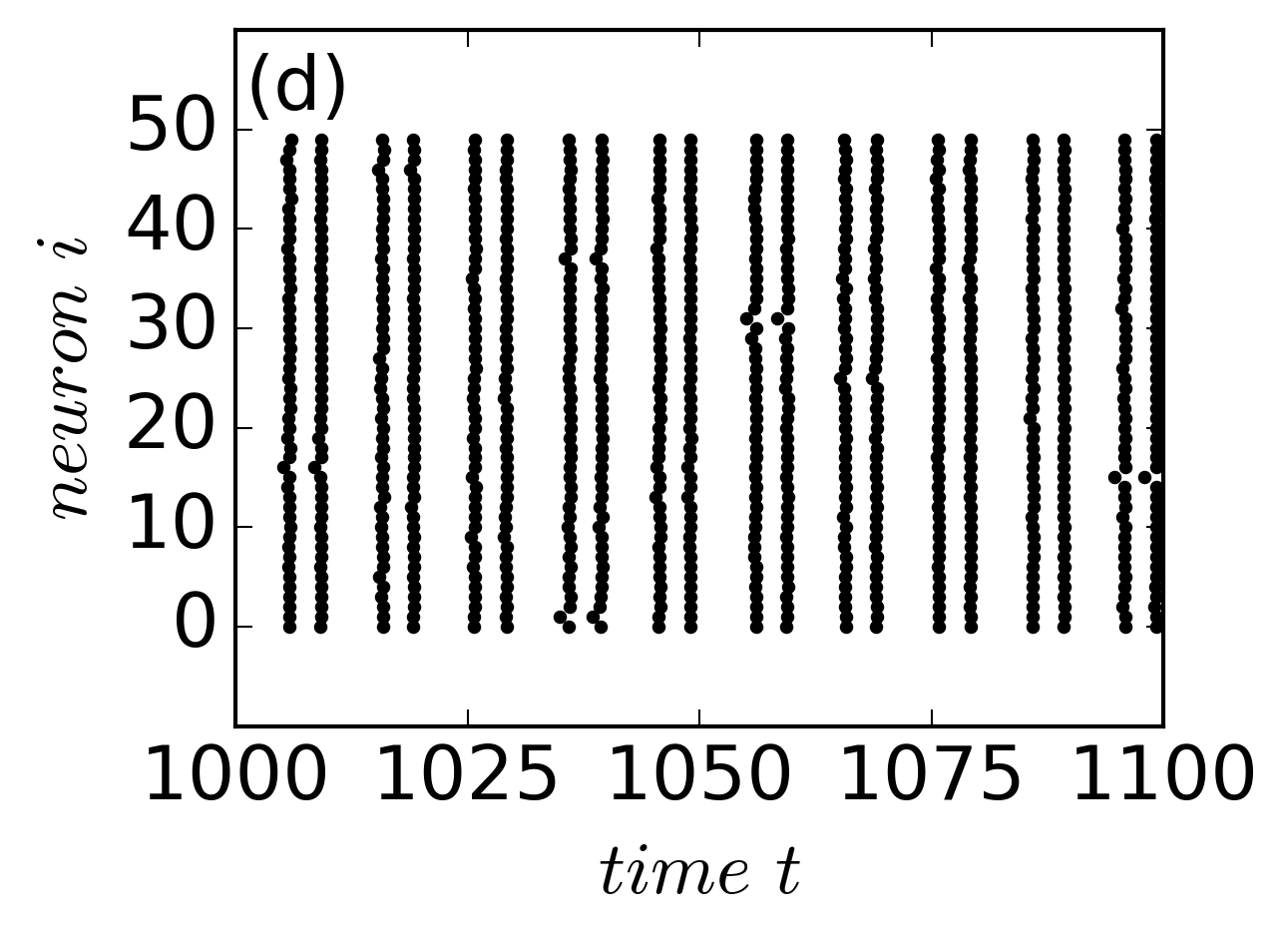}
\end{subfigure}
\caption{Raster plots displaying the spiking activity of the group of 50 neurons all-to-all coupled over time for (a) $\sigma = 0$, (b) $\sigma = 0.01$, (c) $\sigma = 0.015$ and (d) $\sigma = 0.03$. In all the panels: $D = 2.5\cdot 10^{-6}$, $T = 10$ and $a_0 = 0.1$.}
\label{Fig7}
\end{figure}

In Fig. \ref{Fig4}(c) and \ref{Fig4}(d) we note that for the lower signal amplitude, the probabilities vary gradually when increasing the number of links, and with just few links ($\sim$ 10 $\%$) they take the most extreme values, i.e., the encoding is optimal. In contrast, for the higher signal amplitude the probabilities increase or decrease fast, and then saturate. Next, in Fig. \ref{Fig4}(e) and \ref{Fig4}(f) we evaluate the effect of the coupling strength. We notice that increasing $\sigma$ tends to improve the encoding of the signal (the ordinal probabilities tend to higher or lower values), and the effect is more pronounced if the signal amplitude is high. We also note a saturation effect, as for the high signal amplitude, patterns 012 and 210 have zero probability for coupling strengths above $\sigma = 0.02$. 
In order to understand the effect of the coupling strength, Fig.~\ref{Fig7} displays the spiking activity of the neurons for different values of $\sigma$. Here we see that when the neurons are uncopled ($\sigma=0$) their spiking activity is partially synchronized due to the periodic signal that is perceived by all the neurons. As $\sigma$ increases, the spikes gradually become even more synchronized. A similar behavior is found (not shown) when the number of existing links increases, keeping $\sigma$ constant. For future work, it will be interesting to investigate the synchronization transition using synchronization measures based on the ordinal probabilities \citep{ECH19}. 

\section{Conclusions and discussion}
\label{Conclusions}
We have analyzed a plausible neuronal mechanism for encoding a weak periodic signal exploiting neural noise.  We have simulated the dynamics of a neuronal ensemble using the stochastic FitzHugh-Nagumo model with mutual gap-junction type of coupling, and a sinusoidal signal that is perceived by all the neurons. We applied the ordinal symbolic method to the spike sequences generated by all the neurons.  Considering the variation of the ordinal probabilities with the amplitude of the signal, we have found that a group of 50 neurons encodes a weak amplitude signal in a similar way (similar probabilities) as two neurons encode a signal of stronger amplitude. We confirmed the results reported in Refs.~\cite{REI16,MAS18}: the ordinal probabilities depend on the period and the amplitude of the signal and thus, they encode the signal information. We have found that the probabilities have resonances with the period or with the noise level, which become more pronounced for the neuronal ensemble. 
Regarding the influence of the number of neurons, $N$, we have found that increasing $N$ enhances the signal encoding, but above a certain $N$ (which depends on the parameters), the ordinal probabilities saturate and remain nearly constant. We have also investigated the role of the number of links and found that signal encoding can be enhanced by just a few links. We have also found a gradual similar effect when increasing the coupling strength. 

In sum, our work concludes that the neuronal ensemble improves signal encoding, in comparison with single or two coupled neurons. We have studied an homogeneous group of neurons as a first step to understand the ensemble coding mechanism. Yet, in real biological organisms signal coding is performed by nonidentical neurons, and for this reason it will be important to understand the effects of heterogeneous parameters.

The ensemble encoding mechanism proposed here can also allow to encode aperiodic signals, whose amplitude and/or period vary in time. If the ordinal probabilities are determined from the spikes of a single neuron, the encoding mechanim is very slow, because a large number of spikes are needed to estimate the ordinal probabilities; in contrast, when the signal is perceived by a large group of neurons and the ordinal probabilities are determined from the spikes of all the neurons, then signal encoding can be fast, because just a few spikes per neuron can be sufficient to estimate the probabilities of the different spike patterns. 

As future work, it will also be interesting to study how a weak signal that is perceived by just one neuron (or by a subset of neurons) propagates on the whole ensemble, which would give information of how the signal is transmitted. As well, we intent to study how an ensemble of neurons may encode two weak signals. A recent experimental study \cite{CAR18} of how neurons encode simultaneous auditory stimuli has found that some neurons fluctuate between firing rates observed for each individual sound. It would be interesting to compare with our synthetic model, to contribute to advance the understanding of how neuronal systems process information of multiple simultaneous stimuli. 

\section{Acknowledgments}
This work was supported in part by Spanish MINECO/FEDER grant FIS2015-66503-C3-2-P141 and ICREA ACADEMIA, Generalitat de Catalunya. 
\clearpage

\end{document}